\documentclass[sigconf]{acmart}

\usepackage{amsmath,amssymb,amsfonts}
\usepackage{comment}
\usepackage{algorithmic}
\usepackage{graphicx}
\usepackage{textcomp}
\usepackage{xcolor}
\usepackage{color}
\usepackage{xspace}
\usepackage{url}
\usepackage{listings}
\usepackage{amsmath}
\usepackage{tabularx}
\usepackage{booktabs}
\usepackage{multirow}
\usepackage{pifont}
\usepackage{hyperref}
\usepackage{cleveref}
\usepackage{subcaption}
\usepackage{microtype}
\usepackage{flushend}

\graphicspath{{pics/}}

\newcommand{\specialcell}[2][c]{\begin{tabular}[#1]{@{}c@{}}#2\end{tabular}}

\definecolor{medblue}{rgb}{0,0,0.5}
\definecolor[named]{NewtonRed1}{cmyk}{0,0.90,0.86,0}
\definecolor[named]{NewtonRed2}{cmyk}{0,0.84,0.76,0.40}
\definecolor{lightergray}{gray}{0.85}
\definecolor[named]{ACMPurple}{cmyk}{0.55,1,0,0.15}
\definecolor{dkgreen}{rgb}{0,0.6,0}
\definecolor{gray}{rgb}{0.5,0.5,0.5}
\definecolor{mauve}{rgb}{0.58,0,0.82}
\definecolor[named]{ACMDarkBlue}{cmyk}{1,0.58,0,0.21}
\definecolor{myred}{rgb}{0.545098,0.10196,0.0549}
\definecolor{gray75}{gray}{0.75}

\let\OldTexttt\texttt
\renewcommand{\texttt}[1]{{\fontseries{b}\selectfont \OldTexttt{#1}}}

\lstdefinelanguage{newton}{
	morekeywords={for,template,typename},
	morekeywords={[2]double,unsigned,int,inline,void},
	sensitive=false,
	morecomment=[l]{//},
	morecomment=[s]{/*}{*/},
	morestring=[b]",
	alsoletter=-,
}

\newcommand{\approachname}{\textsc{vps}\xspace}
\newcommand{\vps}{\textsc{vps}\xspace}

\newcommand{\vtblptr}{\emph{vtblptr}\xspace}
\newcommand{\thisptr}{\emph{thisptr}\xspace}

\newcommand{\ie}{i.e.}
\newcommand{\eg}{e.g.}

\newcommand{\tabh}[1]{{\bfseries #1}}
\newcommand{\tabsh}[1]{{\itshape #1}}
\newcommand{\tabshc}[1]{\#{\itshape \hspace{1pt}#1}}

\newcommand{\xmark}{\ding{55}\xspace}
\newcommand{\cmark}{\ding{51}\xspace}

\def\first{(i)\xspace}
\def\second{(ii)\xspace}
\def\third{(iii)\xspace}
\def\fourth{(iv)\xspace}

\setcopyright{acmlicensed}

\begin{document}
\title{VPS: Excavating High-Level C++ Constructs from \\ Low-Level Binaries to Protect Dynamic Dispatching}

\author{Andre Pawlowski}
\email{andre.pawlowski@rub.de}
\affiliation{%
	\institution{Ruhr-Universit\"at Bochum}
}

\author{Victor van der Veen}
\email{vvdveen@cs.vu.nl}
\affiliation{%
	\institution{Vrije Universiteit Amsterdam}
}

\author{Dennis Andriesse}
\email{da.andriesse@few.vu.nl}
\affiliation{%
	\institution{Vrije Universiteit Amsterdam}
}

\author{Erik van der Kouwe}
\email{e.van.der.kouwe@liacs.leidenuniv.nl}
\affiliation{%
	\institution{Leiden University}
}

\author{Thorsten Holz}
\email{thorsten.holz@rub.de}
\affiliation{%
	\institution{Ruhr-Universit\"at Bochum}
}

\author{Cristiano Giuffrida}
\email{giuffrida@cs.vu.nl}
\affiliation{%
	\institution{Vrije Universiteit Amsterdam}
}

\author{Herbert Bos}
\email{herbertb@cs.vu.nl}
\affiliation{%
	\institution{Vrije Universiteit Amsterdam}
}

\renewcommand{\shortauthors}{Pawlowski et al.}

\copyrightyear{2019}
\acmYear{2019}
\acmConference[ACSAC '19]{2019 Annual Computer Security Applications 
	Conference}{December 9--13, 2019}{San Juan, PR, USA}
\acmBooktitle{2019 Annual Computer Security Applications Conference (ACSAC '19), 
	December 9--13, 2019, San Juan, PR, USA}
\acmPrice{15.00}
\acmDOI{10.1145/3359789.3359797}
\acmISBN{978-1-4503-7628-0/19/12}



\begin{abstract}

  Polymorphism and inheritance make C++ suitable for writing complex software, but significantly increase the
  attack surface because the implementation relies on \emph{virtual function
    tables (vtables)}. These vtables contain function pointers 
	that attackers can potentially hijack and in practice, \emph{vtable hijacking}
    is one of the most important attack vector for C++ binaries.

    In this paper, we present \emph{VTable Pointer Separation} (\approachname),
	a practical binary-level defense against vtable hijacking in C++ applications.
	Unlike previous binary-level defenses, which rely on unsound static analyses
	to match classes to virtual callsites, \approachname achieves a more
	accurate protection by restricting virtual callsites to validly created objects.
	More specifically, \approachname ensures that virtual callsites can only use
	objects created at valid object construction sites, and only if those objects
	can reach the callsite.
	Moreover, \approachname explicitly prevents false positives (falsely identified virtual callsites)
        from breaking the binary, an issue
        existing work does not handle correctly or at all.
	We evaluate the prototype implementation of \approachname on a diverse set of complex, real-world applications
	(MongoDB, MySQL server, Node.js, SPEC CPU2017/CPU2006), showing that our approach protects 
	on average 97.8\% of all virtual callsites in SPEC CPU2006 and 97.4\% in SPEC CPU2017 (all C++ benchmarks),
	with a moderate performance overhead of 11\% and 9\% geomean, respectively.
	Furthermore, our evaluation
	reveals 86 false negatives in \emph{VTV}, a popular source-based defense which is part of GCC.
	
\end{abstract}

\maketitle

\section{Introduction}

Software implemented in the C++ language is vulnerable to increasingly
sophisticated memory corruption attacks~\cite{schuster2015counterfeit, goktas2014out, carlini2015control,
  conti2015losing, memoryerrors}.  C++ is often the language of
choice for complex software because it allows developers to structure
software by encapsulating data and functionality in \emph{classes},
simplifying the development process.  Unfortunately, the binary-level
implementations of C++ features such as polymorphism and inheritance
are vulnerable to control-flow hijacking attacks, most notably
\emph{vtable hijacking}.
This attack technique abuses common binary-level implementations of C++ virtual methods
where every object with virtual methods contains a pointer to
a \emph{virtual function table} (\emph{vtable}) that stores the addresses of all
the class's virtual functions.
To call a virtual function, the compiler inserts an indirect call
through the corresponding vtable entry (a \emph{virtual callsite}).
Using temporal or spatial memory corruption vulnerabilities such as arbitrary write primitives or use-after-free bugs, 
attackers can overwrite the vtable pointer so that subsequent 
virtual calls use addresses in an attacker-controlled
alternative vtable, thereby hijacking the control flow.
In practice, vtable hijacking is a common exploitation
technique widely used in exploits that target complex applications
written in C++ such as web browser and server applications~\cite{tice2012improving}.

Control-Flow Integrity (CFI) solutions~\cite{abadi2005control,
tice2014enforcing, niu2014modular, bletsch2011mitigating, philippaerts2011code}
protect indirect calls by verifying that control flow is consistent
with a Control-Flow Graph (CFG) derived through static analysis.
However, most generic CFI solutions do not take C++ semantics into account
and leave the attacker with enough wiggle room to build an exploit~\cite{schuster2015counterfeit, goktas2014out}.
Consequently, approaches that specifically protect virtual callsites
in C++ programs have become popular.
If source code is available, compiler-level defenses can benefit from the
rich class hierarchy information available at the
source level~\cite{tice2014enforcing, zhang2016vtrust, bounov2016protecting,
burow2018cfixx}.
However, various legacy applications are still in use \cite{gao2016} or
proprietary binaries have to be protected which do not offer access to the source code
(\eg, Adobe Flash \cite{adobeflash2019}).
Here, binary-level defenses~\cite{pawlowski2017marx,
prakash2015vfguard, elsabagh2017strict, gawlik2014towards, zhang2015vtint} must rely on (automated) binary analysis techniques to reconstruct the information needed to guarantee
security and correctness.

In this paper, we present \emph{VTable Pointer Separation} (\approachname), a binary-level
defense against vtable hijacking attacks.
Unlike previous binary-only approaches that restrict the set of vtables permitted for each virtual callsite,
we check that the vtable pointer remains unmodified after object creation.
Intuitively, \approachname checks the vtable pointer's integrity at every callsite.   Because the vtable pointer in a legitimate live object never changes and the virtual callsite uses it
to determine its target function, \approachname effectively prevents vtable hijacking attacks. %
In essence, we want to bring a defense as powerful as \emph{CFIXX}~\cite{burow2018cfixx} (which operates at the source level) to binary-only applications, even though none of the information needed for the defense is available.
Our approach is suitable for binaries because, unlike other binary-level solutions,
we avoid the inherent inaccuracy in binary-level CFG and class hierarchy reconstruction. 
Because \approachname allows only the initial virtual pointer(s) of the object to ever exist,
we reduce the attack surface even compared to hypothetical implementations of prior approaches that statically find the set of possible vcall targets with perfect accuracy.

Given that binary-level static analysis is challenging
and unsound in practice, and may lead to false positives in identifying virtual callsites, we carefully deal with such cases by
over-approxi\-ma\-ting the set of callsites and implementing an (efficient) slow path to handle possible false positives at runtime. Meanwhile, \approachname handles all previously verified callsite with high optimized fast checks.  
This approach allows us to prevent false positives from breaking the application as they do
in existing work~\cite{prakash2015vfguard, elsabagh2017strict, gawlik2014towards, zhang2015vtint}.
Additionally, while existing work~\cite{pawlowski2017marx,
jin2014recovering, katz2016estimating, katz2018statistical}
only considers \emph{directly} referenced vtables, compilers
also generate code that references vtables \emph{indirectly}, \eg, through the Global
Offset Table (GOT). \approachname can find all code locations that
instantiate objects by writing the vtable, including objects with indirect vtable references.

Our prototype of \approachname is precise enough to
handle complex, real-world C++ applications such as MongoDB, MySQL server, Node.js, and all
C++ applications contained in the SPEC CPU2006 and CPU2017 benchmarks.
Compared to the source code based approach \emph{VTV},
which is part of GCC~\cite{tice2014enforcing}, we can
on average correctly identify 97.8\% and 97.4\% of the virtual callsites in
SPEC CPU2006 and SPEC CPU2017,
with a precision of 95.6\% and 91.1\%, respectively.
Interestingly, our evaluation also revealed 86 virtual callsites
that are \emph{not} protected by VTV, even though it has access to the source code.
A further investigation with the help of the VTV maintainer
showed that this is due to a conceptual problem in VTV which requires non-trivial engineering to fix.
Compared to the source code based approach CFIXX,
\approachname shows an accuracy of 99.6\% and 99.5\% on average for
SPEC CPU2006 and CPU2017 with a precision of 97.0\% and 96.9\%.
These comparisons show that \approachname's binary-level protection of virtual callsites 
closely approaches that of source-level solutions.
While this still leaves a small attack window, it further closes the gap
between binary-only and source-level approaches making vtable hijacking attempts mostly impractical.

Compared to state-of-the-art binary-level analysis frameworks like \emph{Marx}~\cite{pawlowski2017marx},
our analysis identifies
26.5\% more virtual callsites in SPEC CPU2017 and thus offers improved protection.
\approachname induces
geomean performance overhead of 9\% for all C++ applications in SPEC CPU2017 and 11\% for SPEC CPU2006,
which is slightly more than \emph{Marx} induces but with significantly better protection.

\smallskip
\noindent\textbf{Contributions}.
We provide the following contributions:

\begin{itemize}
	\item We present \approachname, a binary-only defense against vtable hijacking attacks
	that sidesteps the imprecision problems of prior work on this topic. The key insight is that vtable pointers only
	change during initialization and destruction of an object (never in between), a property that \approachname can efficiently enforce.
	\item We develop an instrumentation approach that is capable of handling
	false positives in the identification of C++ virtual callsites which would
	otherwise break the application and which  most existing work ignores. Unlike prior work, we also handle indirect vtable references.
	\item Our evaluation shows that our binary-level instrumentation protects nearly the same number of virtual callsites
	as the source-level defenses \emph{VTV} and \emph{CFIXX}.
    In addition, our evaluation uncovered a conceptual problem causing false negatives in VTV (part of gcc).
\end{itemize}

We will release our \approachname implementation and the data we used for the evaluation as open source once this paper is published.

\section{C++ at the Binary Level}

This section provides background on C++ internals needed to understand how \approachname handles C++ binaries.
We focus on how high-level C++ constructs translate to the binary level.
For a more detailed overview of high-level C++ concepts, we refer to the corresponding literature~\cite{anderson1992c++}.

\subsection{Virtual Function Tables}
\label{sec:background_vtable}

C++ supports the paradigm of object-oriented programming (OOP) with polymorphism and (multiple) inheritance.
A class can inherit functions and fields from
another class. The class that inherits is called
the \emph{derived} class and the class from which it inherits is the \emph{base} class.
In addition to \emph{single inheritance} (one class inherits from one other class),
C++ also allows \emph{multiple inheritance}, where 
a derived class has multiple base classes.
A base class can declare a function as \emph{virtual}, which
allows derived classes to override it with their own implementations.
Programmers may choose not to implement some functions in a base class, so called \emph{pure virtual} functions.
Classes containing such functions are \emph{abstract} classes
and cannot be instantiated. Classes deriving from an abstract base
can only be instantiated if they override all pure virtual functions.

Polymorphism is implemented at the binary level using
\emph{virtual function tables} (\emph{vtables}) that
consist of the addresses of all
virtual functions of a particular class.
Each class containing at least one virtual function has a vtable.
Instantiated classes (called \emph{objects})
hold a pointer to their corresponding vtable, which is typically stored in read-only memory.
Since each class has its own corresponding vtable, it can also be considered as
the type of the object.
Throughout this paper,
we refer to the pointer to a vtable as a \vtblptr, while
the pointer to the object is called \thisptr.

\begin{figure}[!ht]
	\centering
	\includegraphics[width=.45\textwidth]{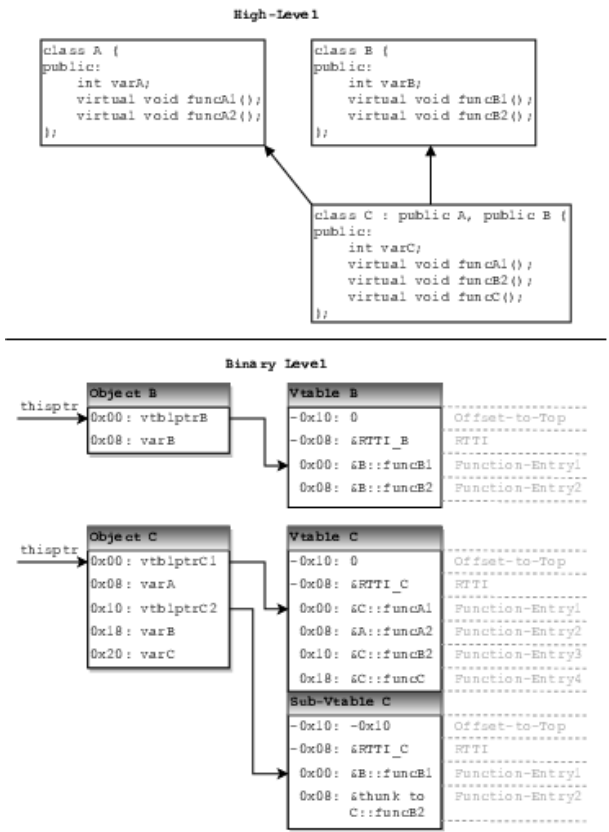}
	\caption{Example C++ class structure. The code at the top
		shows base classes \emph{A} and \emph{B}; derived
		class \emph{C} which overrides virtual functions
		\emph{funcA1} and \emph{funcB2}.
		The bottom shows the binary-level structure of objects of
		classes \emph{B} and \emph{C}.}
	\label{fig:classes}
\end{figure}

The Itanium C++ ABI~\cite{itaniumcppabi} defines the vtable layout for Linux systems.\footnote{Linux uses the Itanium C++ ABI for x86-64 (amd64), our target architecture.}
The \vtblptr points to the first function entry in the vtable, and
the vtable contains an entry for each virtual function
(either inherited or newly declared) in the class.
For example, in Figure~\ref{fig:classes}, class \emph{B}'s
vtable contains two function entries because the class
implements virtual functions \emph{funcB1} and \emph{funcB2}.
Class \emph{C} inherits from two classes, \emph{A} and \emph{B}, and therefore has
two vtables (a base vtable and one sub-vtable). The base vtable
contains all virtual functions inherited from class \emph{A} and implemented by class \emph{C}.
The sub-vtable is a copy of class \emph{B}'s vtable with a special entry that refers
to the overwritten virtual function (called a \emph{thunk}).
Preceding the function entries, a vtable has two metadata fields:
\emph{Runtime Type Identification} (RTTI) and \emph{Offset-to-Top}.
RTTI holds a pointer to type information about the class.
Among other things, this type information contains the name
of the class and its base classes.
However, RTTI is optional and often omitted by the compiler.
It is only needed when the programmer uses, \eg, \emph{dynamic\_cast}
or \emph{type\_info}. Hence, a reliable static analysis cannot rely
on this information. Classes that do not contain RTTI have the RTTI field set to zero.
Offset-to-Top is needed when a class uses multiple inheritance (hence has a base vtable
and one or more sub-vtables) as class \emph{C} does.
Offset-to-Top specifies the distance between a sub-vtable's own \vtblptr and 
the base \vtblptr at the beginning of the object. In our example, 
the \vtblptr to class \emph{C}'s sub-vtable resides at offset \texttt{0x10} in the object, while the
\vtblptr to the base vtable resides at offset \texttt{0x0}.
Hence, the distance between the two, as stored in the Offset-to-Top field in sub-vtable \emph{C}, is \texttt{-0x10}.
Offset-to-Top is \texttt{0} if the vtable is the base vtable of the class
or no multiple inheritance is used.

Vtables can contain one additional field, called \emph{Virtual-Base-Offset},
but it is only used in case of virtual inheritance, an advanced C++ feature
for classes that inherit from the same base multiple times
(diamond-shaped inheritance). An explanation is out of scope here
because \approachname needs no adaptations to support virtual inheritance,
so we defer to the ABI~\cite{itaniumcppabi}.

\subsection{C++ Object Initialization and Destruction}

Because \approachname secures virtual callsites by protecting the \vtblptr
set at initialization time, we explain object initialization
of classes with vtables. 
For the remainder of this paper, we only consider classes
and objects that have at least one virtual function and therefore a vtable.

During object instantiation, the \vtblptr is written into the object by the \emph{constructor}.
The lower part of Figure~\ref{fig:classes} depicts an object's memory layout at the binary level.
The \vtblptr is at offset \texttt{0x0}, the start of the object.
For classes with multiple inheritance, the constructor also initializes
\vtblptr{}s to the sub-vtable(s). In addition, the programmer may initialize
class-specific fields in the constructor.
These fields are located after the \vtblptr and, in case of multiple inheritance,
after any sub-\vtblptr{}s.

For classes that have one or more base classes, the constructors
of the base classes are called before the derived class's own initialization code.
As a result, the base class places its \vtblptr into the
object, which is subsequently overwritten by the derived class's \vtblptr.
Depending on the optimization level,
constructors are often inlined, which may complicate binary analysis that aims to detect constructors.

An analogous principle is applied for object destruction through \emph{destructor} functions.
However, the destructors are executed in reversed order (destructor of the base class is executed last).

Abstract classes form a special case: although programmers cannot instantiate
abstract classes, and despite the fact that their vtables contain \emph{pure\_virtual} function entries,
the compiler can still emit code that writes the \vtblptr to an abstract class into an object.
However, this happens only when creating or releasing an object of a derived class, 
and the abstract \vtblptr is immediately overwritten.

\begin{table*}[t]
	\centering
	\caption{C++ binary-only mitigation mechanisms}
	\scalebox{0.75} {
        \begin{tabular}{l c c c c c l }
            \toprule

			\multirow{2}{*}{\tabh{Defense}} &
			\multirow{2}{*}{\tabh{Binary-only}} & %
  
			\multirow{2}{*}{\specialcell{\tabh{Protects}\\\tabh{vcalls}}} & 
			\multirow{2}{*}{\specialcell{\tabh{Protects}\\\tabh{type}}} & 
			\multirow{2}{*}{\specialcell{\tabh{Protects}\\\tabh{dangl. ptrs}}} & 
			\multirow{2}{*}{\specialcell{\tabh{Tolerates}\\\tabh{FP vcalls}}} & 
			\multirow{2}{*}{\tabh{Security Strategy}} \\
			
			\\
			\midrule
					
			\emph{Marx} (VTable)~\cite{pawlowski2017marx} &
			\cmark & %
			\cmark & %
			\xmark & %
			\cmark & %
			\cmark & %
      \vtblptr in reconstructed class hierarchy (fallback \emph{PathArmor}~\cite{van2015practical}). \\ %
			
			\emph{Marx} (Type-safe)~\cite{pawlowski2017marx} &
			\cmark & %
			\xmark & %
			\xmark & %
			\cmark & %
			n.a. & %
			Memory allocator uses class hierarchy as type. \\ %
			
			\emph{vfGuard}~\cite{prakash2015vfguard} &
			\cmark & %
			\cmark & %
			\xmark & %
			\cmark & %
			\xmark & %
			Call target resides in at least one vtable at correct offset. \\ %
			
			\emph{T-VIP}~\cite{gawlik2014towards} &
			\cmark & %
			\cmark & %
			\xmark & %
			\cmark & %
			\xmark & %
			\vtblptr and random vtable entry must point to read-only memory. \\ %
			
			\emph{VTint}~\cite{zhang2015vtint} &
			\cmark & %
			\cmark & %
			\xmark & %
			\cmark & %
			\xmark & %
			Verifies vtable ID, vtable must be in read-only memory. \\ %
			
			\emph{VCI}~\cite{elsabagh2017strict} &
			\cmark & %
			\cmark & %
			\xmark & %
			\cmark & %
			\xmark & %
			\vtblptr must be statically found, in class hierarchy, or \emph{vfGuard}-allowed. \\ %
			
			\emph{VTPin}~\cite{sarbinowski2016vtpin} &
			needs RTTI & %
			\xmark & %
			\xmark & %
			\cmark & %
			n.a. & %
			Overwrites \vtblptr when object freed. \\ %
				
			\midrule
				
			\emph{VPS} &
			\cmark & %
			\cmark & %
			\cmark & %
			\cmark & %
			\cmark & %
			Check at vcall if object was created at a legitimate object creation site. \\ %

			\bottomrule
		\end{tabular}
	}
	\label{tab:approaches}
	\vspace{-1em}	
\end{table*}

\subsection{C++ Virtual Function Dispatch}
\label{sec:background_vcall}

Because classes can override virtual functions, the compiler cannot determine the target of a
call to such a function at compile time. Therefore, the emitted binary code
uses an indirect function call through the vtable of the object.
This is called a \emph{virtual function call}, or \emph{vcall} for short.
In the Itanium C++ ABI~\cite{itaniumcppabi},
the compiler emits the following structure:
\begin{align*}
   &\texttt{mov\ RDI,\ }\thisptr \\
   &\texttt{call\ [}\vtblptr\texttt{\ +\ offset]}
\end{align*}

The \thisptr is an implicit call argument, so it is moved into the first
argument register, which is \texttt{RDI} on Linux x86-64 systems.
Next, the \texttt{call} instruction uses the \vtblptr to fetch the target function address from the object's vtable.
The \texttt{offset} added to the \vtblptr selects the correct vtable entry.
Note that the offset is a constant, so that corresponding virtual function entries
must be at the same offset in all vtables of classes that inherit from the same base class.

The same code structure holds for cases that use multiple inheritance.
Depending on which (sub-)vtable the virtual function entry
resides in, the \vtblptr either points to the base vtable or
one of the sub-vtables.
However, if the \vtblptr points to a sub-vtable,
\thisptr does not point to the beginning of the object, but rather 
to the offset in the object where the used \vtblptr lies.
Consider the example from Figure~\ref{fig:classes}:
when a function in the sub-vtable of class \emph{C} is called, the call uses the \vtblptr
to its sub-vtable, and the \thisptr points to offset \texttt{0x10} of the object.
Because the code structure is the same, the program treats calls through sub-vtables and base vtables as analogous.

\subsection{Threat Model: VTable Hijacking Attacks}

As we explained in Section~\ref{sec:background_vcall}, virtual callsites use the \vtblptr to
extract the pointer to the called virtual function.
Since the object that stores the \vtblptr is dynamically created during runtime 
and resides in writable memory, an attacker can overwrite it and hijack the control flow at a virtual callsite.

The attacker has two options to hijack an object, depending on the available vulnerabilities:
leveraging a vulnerability to overwrite the object
directly in memory, or using a dangling pointer to an already deleted
object by allocating attacker-controlled memory at the same position
(e.g., via a use-after-free vulnerability).
In the first case,
the attacker can directly overwrite the object's \vtblptr
and use it to hijack the control flow at a vcall.
In the second case, the attacker does not need to overwrite any memory; instead, the vulnerability causes a virtual callsite to use a still existing pointer
to a deleted memory object.
The attacker can control the \vtblptr by allocating new memory at the same
address previously occupied by the deleted object.

We assume the attacker has an arbitrary memory read/write
primitive, and that the $W\oplus{}X$ defense is in place as well as the vtables reside in read-only memory.
These are standard assumptions in related work~\cite{abadi2005control,tice2014enforcing,
	elsabagh2017strict, zhang2015vtint}.
The attacker's goal is to hijack the control flow at a virtual callsite (forward control-flow transfer).
Attacks targeting the backward control-flow transfer (\eg, return address overwrites)
can be secured, for example, by shadow stacks which are orthogonal to \approachname and thus out of scope.
Furthermore, data-only attacks are also out of scope.

\subsection{Related Work on Binary-only Defenses}
\label{sec:related-bin}

Here, we briefly compare our design against binary-only related work as shown
in Table~\ref{tab:approaches}. A detailed discussion including source-level
approaches is provided in Section~\ref{sec:related}.
 
Most existing vtable hijacking defenses assign a set of allowed target functions
to each virtual callsite (\eg, \emph{Marx} \emph{VTable Protection}~\cite{pawlowski2017marx},
\emph{vfGuard}~\cite{prakash2015vfguard}, T-VIP~\cite{gawlik2014towards}, \emph{VTint}~\cite{zhang2015vtint} and \emph{VCI}~\cite{elsabagh2017strict}).
The inaccuracy of binary analysis forces them to overestimate
the target set, leaving room for attacks~\cite{schuster2015counterfeit}.
In contrast, \approachname enforces that vtable pointers remain unmodified after object construction,
ensuring that only validly created objects can be used at virtual callsites and reducing the
attack surface even compared to a hypothetical defense with a perfect set of allowed targets.
\emph{Marx} \emph{Type-safe Object Reuse} and \emph{VTPin}~\cite{sarbinowski2016vtpin} protect against
the reuse of dangling pointers by modifying the memory allocator. \approachname protects against dangling pointers \emph{without} any further modification.

As the comparison in Table~\ref{tab:approaches} shows, \approachname combines the protection targets given by related work
and additionally protects the type integrity of the object itself.

\section{VTable Pointer Separation}

\begin{figure*}
	\centering
	\includegraphics[width=.99\textwidth]{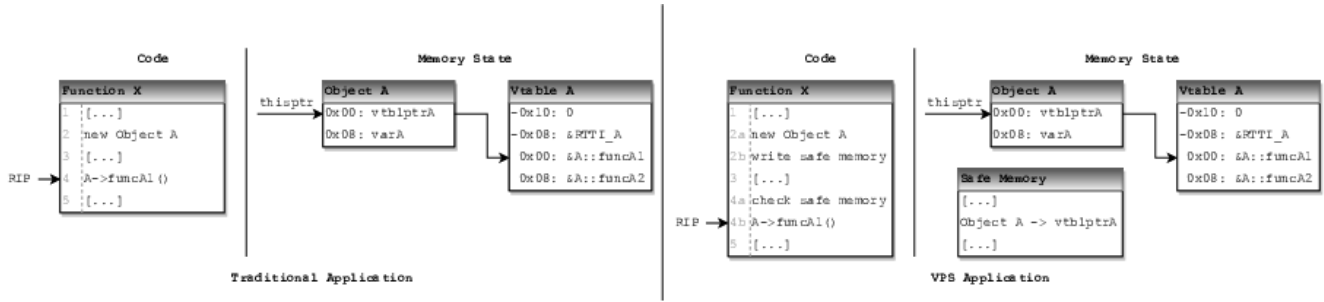}
	\caption{High-level overview of the object instantiation
		and virtual callsite of a traditional application (left side)
		and a \approachname protected application (right side).
		For both applications the memory state is given while the
		instruction pointer executes the function call.}
	\label{fig:overview}
\end{figure*}

Our approach is based on the observation
that the \vtblptr is only written during object
initialization and destruction and cannot legitimately change in between.
Therefore, only the \vtblptr that is written
by the constructor (or destructor) is a valid value.
If a \vtblptr changes between the object was created and destroyed,
a vtable hijacking attack is in progress.
Since these attacks target virtual callsites, it is sufficient
to check at each virtual callsite if the \vtblptr written originally
into the object still resides there.

Figure~\ref{fig:overview} depicts the differences between a traditional
application and a \approachname-protected application.
The traditional application
initializes an object and uses a vcall and the
created object to call a virtual function. As explained
in Section~\ref{sec:background_vcall}, the application uses the vtable
to decide which virtual function to execute. If an attacker is able
to corrupt the object between the initialization and vcall, she can
place her own vtable in memory and hijack the control flow.
In contrast, the \approachname-protected application adds two
additional functionalities to the executed code. While
the object is initialized, it stores the \vtblptr in a
safe memory region. Before a vcall,
it checks if the \vtblptr in the object
is still the same as the one stored for the object in the
safe memory region. The vcall is only executed when the check succeeds.
As a result, the same attacker that is able to corrupt the
object in between can no longer hijack the control flow.
The same concept holds for \vtblptr{}s written in the destructors.
The \vtblptr is written into the object and used for vcalls during its
destruction (if it is used at all). Since a \approachname-protected application
stores the written \vtblptr into the safe memory region and checks
the integrity of the one in the object if it is used at a vcall,
the approach does not need to differentiate between object initialization and destruction.

In contrast to other binary-only defenses for virtual
callsites \cite{gawlik2014towards, prakash2015vfguard,
	zhang2015vtint, pawlowski2017marx, elsabagh2017strict}
that allow a specific overestimated set of classes at a virtual
function dispatch, \approachname has a direct mapping
between an object initialization site and the reachable vcalls.

Even though \approachname looks conceptually similar to CFIXX, adding this protection at the binary level
encounters multiple hurdles. Performing accurate analysis at the binary level is a challenging problem,
especially with regards to object creation sites, where false negatives would break the protected application.
Our analysis has to take direct and indirect vtable accesses into account, which do not exist
on the source level. The virtual callsite identification has to be as precise as possible
in order to provide a high level of security and it has to be performed without type information.
Any false positive in this result breaks the application, which makes an instrumentation capable of
handling these necessary (a problem that other binary-only approaches do not consider).

\section{Analysis Approach}
\label{sec:analysis_approach}

\approachname protects binary C++ applications against control-flow
hijacking attacks at virtual callsites.
To this end, we first analyze the binary to identify C++-specific properties
and then apply instrumentation to harden it.
We divide the analysis into three phases: \emph{Vtable Identification},
\emph{Vtable Pointer Write Operations}, and
\emph{Virtual Callsite Identification}.
At a high-level, our analysis first identifies all vtables in the target binary
in the Vtable Identification phase. Subsequently, the identified vtables are
used to find all locations in the binary that write \vtblptr{}s. Eventually, 
the identified vtables are also used to identify and verify vcalls in the
Virtual Callsite Identification phase.
While the Vtable Identification static analysis
is an improved and more exact version of Pawlowski~et~al.~\cite{pawlowski2017marx}
(finding vtables in \texttt{.bss} and GOT,
considering indirect referencing of vtables),
the other analyses are novel to \approachname.
In the remainder of this section, we explain the details
of our analysis approach.
Note that we focus on Linux x86-64 binaries that use
the Itanium C++ ABI~\cite{itaniumcppabi}. However, 
our analysis approach is conceptually mostly generic and
with additional engineering effort can be applied
to other architectures and ABIs as well.
For architecture-specific steps in our analysis, we describe 
what to modify to port the step to other architectures.

\subsection{Vtable Identification}
\label{sec:vtableidentify}

To protect \vtblptr{}s in objects,
we need to know the location of all vtables in the binary. 
To find these, our static analysis searches through the binary and uses
a set of rules to identify vtables.
Whenever all rules are satisfied, the algorithm identifies a vtable.
As explained earlier, Figure~\ref{fig:classes} shows a typical vtable structure.
The smallest possible vtable in the Itanium C++ ABI~\cite{itaniumcppabi}
consists of three consecutive words
(\emph{Offset-to-Top}, \emph{RTTI}, and \emph{Function-Entry}).
We use the following five rules to determine the beginning of a vtable:

\paragraph{R-1} In principle, our algorithm searches for vtables in read-only sections such as \texttt{.rodata} and \texttt{.data.rel.ro}.
However, there are exceptions to this. If a class has a base
class that resides in another module and the compiler uses copy relocation, the loader will copy the vtable into the \texttt{.bss} section~\cite{ge2017evil}.
Additionally, vtables from other modules can
be referenced through the Global Offset Table (GOT),
\eg, in position-independent code~\cite{elf}.
To handle these cases where the vtable data lies outside the main binary, we parse the binary's dynamic symbol table
and search for vtables that are either copied to the \texttt{.bss} section or referenced through the GOT.
Note that we do not rely on debugging symbols, only on symbols that the loader uses, which cannot be stripped.

\paragraph{R-2} Recall that the \vtblptr points to the first function entry in a class's vtable, and is written into the object at initialization time.
Therefore, our algorithm looks for code patterns that reference this first function entry.
Again, there are special cases to handle.
The compiler sometimes emits code that does not
reference the first function entry of the vtable,
but rather the first metadata field at offset \texttt{-0x10}
(or \texttt{-0x18} if virtual inheritance is used).
This happens for example in position-independent code.
To handle these cases, we additionally look for code patterns that add \texttt{0x10} (or \texttt{0x18})
to the reference before writing the \vtblptr into the object, which is necessary to comply with
the Itanium C++ ABI~\cite{itaniumcppabi}.
Our algorithm also checks for the special case where vtables are referenced through the GOT instead of directly.

\paragraph{R-3}

As depicted in Figure~\ref{fig:classes}, the \emph{Offset-to-Top}
is stored in the first metadata field of the vtable
at offset \texttt{-0x10}. In most cases this field is
\texttt{0}, but when multiple inheritance is used,
this field gives the distance between the base
\vtblptr and the sub-\vtblptr in the object
(see Section~\ref{sec:background_vtable}).
Our algorithm checks the sanity of this value
by allowing a range between \texttt{-0xFFFFFF}
and \texttt{0xFFFFFF}, as proposed by
Prakash~et~al.~\cite{prakash2015vfguard}.

\paragraph{R-4}

The RTTI field at offset \texttt{-0x8} in the vtable, which can hold a pointer to RTTI metadata, is optional and usually omitted by the compiler.
If omitted, this field holds \texttt{0}; otherwise, it holds a pointer into the \texttt{data} section or a relocation entry if the class inherits from another class in a shared object.

\paragraph{R-5}

Most of the vtable consists of function entries that hold 
pointers to virtual functions. Our algorithm deems
them valid if they point into any of the \texttt{.text},
\texttt{.plt}, or \texttt{.extern} sections of the binary,
or are a relocation entry.

\smallskip
Abstract classes are an edge case. For each
virtual function without implementation, the vtable
points to a special function called \emph{pure\_virtual}.
Because abstract classes are not meant to be instantiated, 
calling \emph{pure\_virtual} throws an exception.
Additionally, the first function entries in a vtable can be \texttt{0} if the compiler did not
emit the code of the corresponding functions (\eg, for destructor functions).
To cope with this, Pawlowski~et~al.~\cite{pawlowski2017marx}
allow \texttt{0} entries in the beginning
of a vtable. 
We omit this rule because our approach can safely ignore the instantiation of
abstract classes, given that \vtblptr{}s for abstract classes are
overwritten shortly after object initialization.

In case of multiple inheritance, we do not
distinguish between vtables and sub-vtables.
That is, in the example in Figure~\ref{fig:classes},
our approach identifies \emph{Vtable C} and \emph{Sub-Vtable C}
as separate vtables. 
As discussed later, this does not pose any limitations for our approach given our focus on \vtblptr write operations (as opposed to methods that couple class hierarchies to virtual call sites).

The combination of multiple inheritance and copy relocation
poses another edge case. In copy relocation, the loader copies 
data residing at the position given by a relocation symbol
into the \texttt{.bss} section without regards to the type of the data.
For classes that use multiple inheritance,
the copied data contains a base vtable and sub-vtable(s), but the corresponding relocation symbol holds only information on the beginning and length of the data, not the vtable locations.
To ensure that we do not miss any, we 
identify every \texttt{8}-byte aligned address
of the copied data as a vtable. For example,
if the loader copies a data chunk of \texttt{0x40} bytes to the address \texttt{0x100}, we identify
the addresses \texttt{0x100}, \texttt{0x108}, \texttt{0x110}, \ldots\,up to \texttt{0x138} as vtables.
While this overestimates the set of vtables, only the correct vtables and sub-vtables are referenced during object initialization.

Note that on other architectures, the assumed size of 8-byte per vtable entry as used
by our rules may have to be adjusted.
For example, Linux on x86 (32-bit) and ARM would use 4-byte entries, with no conceptual changes.

\subsection{Vtable Pointer Write Operations}
\label{sec:object_init_ops}

The next phase of our static analysis is based on
the observation that to create a new object,
its \vtblptr has to be written into the corresponding
memory object during the initialization. This is done in the constructor of
the class which can be either an explicit function or inlined code.
The same holds for object destruction by the corresponding destructor function.
Hence, the goal of this analysis step is to identify the exact
instruction that writes the \vtblptr into the memory object.
This step is Linux-specific but architecture-agnostic.

First, we search for all references from code to the vtables identified in the previous step.
Because vtables are not always referenced directly,
the analysis searches for the following different reference methods:

\begin{enumerate}
	\item A direct reference to the start of the function entries
		  in the vtable. This is the most common case.
	\item A reference to the beginning of the metadata fields
		  in the vtable. This is mostly used by applications
		  compiled with position-independent code (\eg, 
		  MySQL server which additionally uses virtual inheritance).
	\item An indirect reference through the GOT. Here, the
		  address to the vtable is loaded from the GOT.
\end{enumerate}

Starting from the identified references,
we track the data flow through the code (using Static Single Assignment (SSA) form~\cite{cytron1991efficiently}) to the instructions that write the \vtblptr{}s during object initialization or destruction.
We later instrument these instructions, adding code that stores the \vtblptr in a safe memory region.
Our approach is agnostic to the location the C++ object resides in (\ie, heap, stack, or global memory).
Furthermore, since we focus on references from code to the vtables, our approach can handle explicit
constructor functions as well as inlined constructors and destructors.

During our research, we encountered functions
with inlined constructors where the compiler emits
code that stores the \vtblptr temporarily
in a stack variable to use it at
multiple places in the same function.
Therefore, to ensure that we do not miss any \vtblptr write instructions,
our algorithm continues to track the data flow
even after a \vtblptr is written into a stack variable.
Because we cannot easily distinguish between
a temporary stack variable and an object residing
on the stack, our algorithm also assumes that the temporary
stack variable is a C++ object. While this overestimates the set of C++ objects, it ensures that we instrument all \vtblptr write instructions, making this overapproximation comprehensive.

\subsection{Virtual Callsite Identification}

Because \approachname specifically protects vcalls against control-flow
hijacking, we first have to locate them in the target binary. Hence, we 
have to differentiate between vcalls and normal C-style indirect call instructions.
We follow a two-stage approach to make this distinction: we first locate
all possible vcall candidates and subsequently verify them.
The verification step consists of a static analysis component
and a dynamic one.
In the following, we explain this analysis in detail.

\subsubsection{Virtual Callsite Candidates}
\label{sec:vcall_candidates}

To find virtual callsite candidates,
we use a similar technique as
previous work~\cite{prakash2015vfguard, elsabagh2017strict, gawlik2014towards, zhang2015vtint}.
We search for the vcall pattern described in Section~\ref{sec:background_vcall}, where the \thisptr is the first argument (stored in the \texttt{RDI} register on Linux x86-64) to the called function and the
vcall uses the \vtblptr to retrieve the call target from the vtable.
Note that the \thisptr is also used to extract the \vtblptr for the call instruction. A typical vcall looks as follows:
\begin{align*}
   &\texttt{mov\ RDI,\ } \thisptr \\
   &\texttt{mov\ }\vtblptr\texttt{,\ [}\thisptr\texttt{]} \\
   &\texttt{call\ [}\vtblptr\texttt{\ +\ offset]}
\end{align*}

Note that these instructions do not have to be consecutive in
the application, but can be interspersed with other instructions.
Two patterns can be derived from this sequence: the first argument
register always holds the \thisptr, and the call instruction
target can be denoted as \texttt{[[}\thisptr\texttt{]~+~offset]}, where
\texttt{offset} can be \texttt{0} and therefore omitted.
This specific dependency between call target and first argument register
is rare for non-C++ indirect calls.
With the help of the SSA form, our algorithm traces the data flow of the function.
If the previously described dependency is satisfied, we consider the indirect call instruction
a \emph{vcall candidate}.

Note that the same pattern holds for classes with multiple inheritance.
As described in Section~\ref{sec:background_vcall}, when a virtual
function of a sub-vtable is called, the \thisptr
is moved to the position in the object where the sub-vtable resides.
Therefore, the first argument holds \thisptr\texttt{~+~distance},
and the call target \texttt{[[}\thisptr\texttt{~+~distance] + offset]}.
This still satisfies the aforementioned dependency between first argument
and call target. Furthermore, the pattern also applies to
Linux ARM, Linux x86, and Windows x86-64
binaries, requiring only a minor modification to account for the specific register or memory location
used for the first argument on the platform (\texttt{R0} for ARM,
the first stack argument for Linux x86, and \texttt{RCX} for Windows x86-64).

To effectively protect vcalls, it is crucial to prevent false positive vcall identifications, as these may break the application
during instrumentation. This is also required for related work~\cite{prakash2015vfguard, elsabagh2017strict, gawlik2014towards, zhang2015vtint}.
While the authors of prior approaches 
report no false positives with the above vcall identification approach,
our research shows that most larger binary programs
do indeed contain patterns that result in indirect calls being wrongly
classified as virtual callsites.

A possible explanation for the lack of false positives in previous work is that most prior work focuses on Windows x86~\cite{prakash2015vfguard, gawlik2014towards, zhang2015vtint}, where 
the calling conventions for vcalls and other call instructions
differ. That is, on Windows x86, the \thisptr is passed to the virtual function
via the \texttt{ECX} register (\emph{thiscall} calling convention),
while other call instructions pass the first argument
via the stack (\emph{stdcall} calling convention)~\cite{callingconventions}.
This is not the case for Windows x86-64 and Linux (x86 and x86-64).
On these architectures, the \thisptr is passed as the first argument
in the platform's standard calling convention (\emph{Microsoft x64},
\emph{cdecl} and \emph{System V AMD64 ABI}, respectively).
While Elsabagh~et~al.~\cite{elsabagh2017strict},
who work on Linux x86, did not report
false positives, our evaluation does show false positives in the same application set.
We contacted the authors, but they could not help us find an explanation
for these differing outcomes and could not give us access to the source code to allow us to reproduce the results.

\subsubsection{Virtual Callsite Verification}
\label{sec:vcall_verification}

Because a single false positive can break our approach,
the next phase in our static analysis verifies 
the virtual callsite candidates. Basically, we perform a data-flow analysis
in which we track whether a \vtblptr is used at a virtual callsite candidate.
If the candidate uses the \vtblptr to determine the call target, we consider it as verified.
However, a data-flow graph alone is not sufficient to verify this connection. The control flow
and actual usage of the \vtblptr have also to be considered.
The following describes our analysis in detail.

\paragraph{Data-Flow Graphs}

First, our analysis tracks the data flow backwards with the help of SSA form
starting from all vtable references in the code (which create the \vtblptr).
The data flow is tracked over function boundaries
when argument registers or the return value register \texttt{RAX} are involved.
This means the tracking is done interprocedurally. 
The same data-flow tracking is done for the call target of each virtual callsite candidate.
As a result, we obtain data-flow graphs showing the source of the data used by the
vtable-referencing instructions and the virtual callsite candidates.
Whenever a data-flow graph for a virtual callsite candidate has the same data source
as a vtable-referencing instruction, we group them together.

\paragraph{Control-Flow Path}

Virtual callsite candidates and vtable-referen\-cing instructions that share the same data
source represent a possible connection between a created \vtblptr and a corresponding vcall.
However, this connection alone does not give any information on whether the \vtblptr is actually
used at the virtual callsite candidate. To verify this, we have to check if 
a control-flow path exists that starts at the data source instruction,
visits the vtable-referencing instruction, and ends at the vcall instruction.
For this, our analysis searches all possible data-flow paths through the graph
that start at a data source instruction and end in a vtable-referencing instruction, and
that start from a data source instruction and end at a virtual callsite candidate.

Next, our analysis tries to transform these data-flow paths into a control-flow path by translating each data-flow node into the basic block
that contains the corresponding instruction. With the help of the Control-Flow Graph (CFG),
our analysis then searches for a path from basic block to basic block until it reaches the final block.
Eventually, if a path exists, the algorithm finds a possible control-flow path that starts from the data source instruction,
visits the vtable-referencing instruction, and ends at the vcall instruction.

\paragraph{Symbolic Execution}

As a last step, we symbolically execute the obtained control-flow paths to track
the flow of the \vtblptr through the binary.
When an instruction writes a vtable into the memory state,
we replace that \vtblptr with a symbolic value.
To keep the analysis scalable to large real-world applications,
our symbolic execution simply executes basic blocks without checking whether branches can actually be taken in a concrete execution.
If a basic block contains a call instruction that is not part of our original
data-flow path, we simply execute a return instruction immediately after the call instead of symbolically executing the called function.
When the symbolic execution reaches the vcall instruction, we check the obtained memory
state to verify that the \vtblptr is used for the call target. If so,
we conclude that the vcall candidate is in fact a vcall and consider it a \emph{verified vcall}.
\medskip

In addition to explicit vtable-referencing instructions, this analysis phase
checks implicit vtable references as well. In case the earlier backward data-flow analysis
shows that a vcall target stems from the first argument register, we check
whether the calling function is a known virtual function (by checking whether the function resides in
any previously identified vtable). If it is, we add a special virtual function node
to the data-flow graph. We then search for a path from this virtual function
node to the vcall instruction. If a path is found, we apply the steps
described previously for transforming the data-flow path to a
control-flow path. For such paths, before starting the symbolic execution,
we add an artificial memory object containing the \vtblptr and place the \thisptr in the first argument register.
This way, we simulate an implicit use of the vtable through the initialized object.

We perform the whole vcall verification analysis in an iterative manner.
When the data-flow tracking step stops at an indirect call instruction,
we repeat it as soon as our analysis has verified the indirect call as a vcall
and has therefore found corresponding vtables for resolving the target.
The same applies to data-flow tracking that stops at the
beginning of a virtual function (because no caller is known). As soon as we can determine a corresponding
vcall instruction, we repeat the analysis.
The analysis continues until we reach a fixed point where the analysis fails to find any new results.

\subsubsection{Dynamic Profiling}
\label{sec:dynamic_profiling}

Our approach
includes a dynamic profiling phase that further refines the vcall verification. During this phase,
we execute the application with instrumentation code added to all virtual callsite candidates
(only the vcall candidates, not the already verified vcalls).
Whenever the execution reaches a vcall, the instrumentation code verifies that
the first argument contains a valid \thisptr. To verify this, we
check if the first element of the object the \thisptr points to contains a valid pointer
to a known vtable (\vtblptr). If it does, we consider the vcall verified.
Otherwise, we regard the vcall as a false positive of the static analysis and discard it.

Because this phase only instruments vcall candidates identified by
the static analysis described in Section~\ref{sec:vcall_candidates},
it is safe to assume the dependency between first argument and call instruction target.
Hence, the above dynamic profiling check is sufficient to remove
false positives seen during the profiling run, given that
the odds of finding a C-style indirect callsite with such a distinctive pattern that
uses C++ objects is extremely unlikely. We did not encounter any such case
during our comprehensive evaluation.
Also note, that only this dynamic analysis step discards vcall candidates as false positives.
Vcalls that could not be verified by the static analysis (or not reached during this dynamic profiling)
are still considered vcall candidates
since the reason for the failed verification can be missing information (\eg, analysis gaps through
indirect control-flow transfers).

\section{Instrumentation Approach}
\label{sec:instrumentation_approach}

\approachname protects virtual callsites
against control-flow hijacking attacks by instrumenting
the application using the results from the analysis phase.
We instrument two parts of the program:
\emph{Object Initialization and Destruction} and \emph{Virtual Callsites}.
The following describes how both kinds of instrumentation work.

\subsection{Object Initialization and Destruction}

We use
the data collected in Section~\ref{sec:object_init_ops} to instrument object initialization,
specifically the instruction that writes the
\vtblptr into the object.
When an object is created, the instrumentation code stores a key-value pair that uses the memory address
of the object as the \emph{key} and maps it to the \vtblptr, which is the
associated \emph{value}.
To prevent tampering with this mapping, we store it in a safe memory region.

Recall that when a C++ object is created that inherits from another class,
the initialization code first writes the \vtblptr of the base class into the object,
which is then overwritten by the \vtblptr of the
derived class. Our approach is agnostic to inheritance
and simply overwrites the \vtblptr in the same order (because
each \vtblptr write instruction is instrumented).

Similarly, our approach is agnostic to multiple inheritance,
because object initialization sites use the address where the \vtblptr is written
as the object address. As explained in Section~\ref{sec:background_vcall},
at a virtual callsite the \thisptr points to the address of the
object the used \vtblptr resides in. For a sub-vtable, this
is not the beginning of the object, but an offset somewhere
in the object (in our running example in Figure~\ref{fig:classes}
offset \texttt{0x10}). Because this is exactly the address that our approach
uses as the key for the safe memory region, our approach works for
multiple inheritance without any special handling.

Since this instrumentation only focuses on \vtblptr write instructions, it is also agnostic
to object initialization and destruction. Hence, we do not have to differentiate between
constructor and destructor and can use it for both.

Moreover, despite the fact that we ignore
object deletion, our approach does not suffer from consistency problems.
This is because, when an object is deleted and its released memory is reused
for a new C++ object, the instrumentation code for the initialization of this new object
automatically overwrites the old value in the safe memory
region with the current \vtblptr.

\subsection{Virtual Callsites}

Because a single false positive virtual callsite 
can break the application, we designed the vcall instrumentation code
such that it can detect false positives and filter them out. In doing so,
the vcall instrumentation continuously refines the previous analysis results.
The vcall instrumentation consists of two components, described next:
\emph{Analysis Instrumentation} and \emph{Security Instrumentation}.

\subsubsection{Analysis Instrumentation}

We add analysis instrumentation code to all vcall candidates
that we were unable to verify during our static vcall verification and dynamic profiling analysis.
For verified vcall sites, we only add security instrumentation and omit the analysis code.

Before executing a vcall candidate, the analysis instrumentation performs
the same check as the dynamic profiling
phase described in Section~\ref{sec:dynamic_profiling}.
If the check fails, meaning that this is not a vcall but a regular C-style
indirect call, we remove all instrumentation from the call site.
If the check succeeds, we replace the analysis instrumentation
with the more lightweight security instrumentation for verified virtual
callsites described in Section~\ref{sec:security_instrumentation}, and immediately
run the security instrumentation code.

Through our use of adaptive instrumentation, our approach is
able to cope with false positives and further refine the analysis
results during runtime. By caching the refined results on disk,
we can reuse these in later runs of the same application, improving \approachname's performance over time.
Furthermore, caching also improves the security of our adaptive instrumentation as we discuss in Section~\ref{sec:limitations}.

Because the analysis instrumentation verifies all remaining vcall candidates for false positives at runtime, 
the static vcall verification from Section~\ref{sec:vcall_verification}
and the dynamic profiling from Section~\ref{sec:dynamic_profiling}
can be omitted. Omitting these steps does not affect the correctness of our approach,
although we recommend using them for optimal performance.

\subsubsection{Security Instrumentation}
\label{sec:security_instrumentation}

We protect verified vcall sites against control-flow hijacking by adding security instrumentation code that runs before allowing the vcall.
The instrumentation uses
the \thisptr in the first argument register to retrieve the \vtblptr
stored for this object in the safe memory region.
To decide whether to allow the vcall, the instrumentation code compares the \vtblptr from the safe memory region with the
one stored in the actual object used in the vcall.
If they are the same, the instrumentation allows the vcall.
If not, we terminate with an alert.

\section{Implementation}
\label{sec:implementation}

Based on the approach from Section~\ref{sec:analysis_approach},
we integrated our static analysis into the open source \emph{Marx} framework~\cite{pawlowski2017marx}.
This framework provides a basic symbolic execution based
on the VEX-IR from the Valgrind project~\cite{valgrind} and
data structures needed for C++ binary analysis.
It is written in C++ and targets Linux x86-64 (amd64) binaries.
To support integration of our approach into the \emph{Marx} framework,
we added support for SSA and a generic data-flow tracking algorithm.

Because the VEX-IR supports multiple architectures, the framework
is easily extendable to these.
The same is true for our approach, which is mostly independent from the underlying architecture (Section~\ref{sec:analysis_approach}).
To balance precision and scalability, the symbolic execution
emulates only a subset of the 64-bit VEX instructions that
suits our focus on vtable-centered data-flow tracking in real-world applications. 

We use IDAPython~\cite{idapython} for vtable identification and CFG extraction.
Additionally,
we use instruction data provided by IDA Pro to support the SSA transformation, and use
Protocol Buffers~\cite{protobuf} to export the results in a programming language--agnostic format.
We implement dynamic profiling with Pin~\cite{luk2005pin}.
We build the runtime component of \approachname on top of Dyninst v9.3.2~\cite{bernat2011anywhere}.
Dyninst is responsible for
installing \vtblptr write and (candidate) virtual callsite hooks. We
inject these wrappers into the target program's address space by preloading a
shared library. 

To set up the safe memory region, our preloaded library maps the lower half of the address
space as a safe region at load time; this is straightforward for position-independent
executables as their segments are mapped exclusively in the upper half of the
address space by default. To compute safe addresses, we subtract 64\,TB\footnote{Linux x86-64 provides 47 bits for user space mappings, and $2^{47}=$~128\,TB.}
from the addresses used by \vtblptr writes or virtual calls.
To thwart value probing attacks in the safe region, we \first mark all safe region pages as inaccessible by default and make them accessible on demand, 
and \second use a fixed offset chosen randomly at load time for writes to the safe region.
To achieve the latter, we write a random value to the \texttt{gs} register and use it
as the offset for all accesses to the safe region. To mark pages as readable/writable on demand,
we use a custom segfault handler that uses \texttt{mprotect} to allow accesses from our library.
This means that when a \vtblptr is written into the safe memory region and the page is not yet accessible, our segfault
handler checks if the write access is done by our library and makes the page accessible if it is. Otherwise, a probing attack is
detected and execution is stopped. The page remains accessible which speeds up further \vtblptr writes to it.

We omit an evaluation of potential optimizations already explored in prior work~\cite{kuznetsov2014code, burow2018cfixx},
such as avoiding Dyninst's penalties for (re)storing unclobbered live registers or  
removing trampoline code left over after nopping out analysis instrumentation code.
Similarly, we do not implement hash-based safe region compression that would reduce virtual and
physical memory usage and allow increased entropy in the safe region, nor do we use
Intel MPK~\cite{mpk} to further secure the safe region.
Since we focus on the exact analysis of binary applications and the subsequent instrumentation, we consider these optimizations orthogonal to our work.

\section{Evaluation}
\label{sec:eval}

In this section, we evaluate \approachname in terms of performance and
accuracy. We focus our evaluation on MySQL, Node.js, MongoDB, and the fifteen C++ benchmarks found in SPEC CPU2006 and CPU2017~\cite{spec2006, spec2017}.
Even though our approach is able to handle proprietary software, we evaluate it on open source software
since otherwise we are not able to generate a ground truth to compare against.

\subsection{Virtual Callsite Identification Accuracy}
\label{sec:evalvcalls}

\begin{table*}
	\centering
	\caption{Results of our vcall accuracy evaluation. For each application this table
		shows (i) the code size, time needed for the static analysis (hh:mm:ss) and the ground truth generated by VTV;
		(ii) static vcall identification,
		depicting the number of indirect call instructions identified as vcall that
		are true positives, the
		false positives, recall and precision; (iii) static vcall verification results,
		listing the number of verified vcall instructions, verified vcalls in percent
		and verified false positives;
		(iv) static and dynamic verification results, showing the number of verified
		vcall instructions, verified vcalls in percent,
		verified false positives and false positive identified
		vcalls removed. Cases where dynamic verification failed
		due to VTV false positives are in parentheses.}
	\scalebox{0.90} {
		\begin{tabular}{l r r r r r r r r r r r r r r }
			\toprule
			
			&
			&
			&
			& \multicolumn{4}{c}{\tabh{Static Identification}}
			& \multicolumn{3}{c}{\tabh{Static Verification}}
			& \multicolumn{4}{c}{\tabh{Static and Dynamic Verification}} \\
			
			\cmidrule(lr){5-8}
			\cmidrule(lr){9-11}
			\cmidrule(lr){12-15}
			
			\tabh{Program} &
			\tabsh{Code Size} &
			\tabsh{Time} &
			\tabshc{GT} &
			\tabshc{TP} &
			\tabshc{FP} &
			\tabsh{Recall (\%)} &
			\tabsh{Precision (\%)} &
			\tabshc{} &
			\tabsh{\%} &
			\tabshc{FP} &
			\tabshc{} &
			\tabsh{\%} &
			\tabshc{FP} &
			\tabshc{removed}
			\\ 
			
			\midrule
			
			447.dealII &
			4.18 MB & %
			0:02:15 & %
			1,558 & %
			1,450 & %
			215 & %
			93.0 & %
			87.1 & %
			379 & %
			24.3 & %
			7 & %
			423 & %
			27.2 & %
			18 & %
			0 \\ %
			
			450.soplex &
			-- & %
			-- & %
			-- & %
			-- & %
			-- & %
			-- & %
			-- & %
			-- & %
			-- & %
			-- & %
			-- & %
			-- & %
			-- & %
			-- \\ %
			
			453.povray &
			1.09 MB & %
			0:00:04 & %
			102 & %
			102 & %
			10 & %
			100.0 & %
			91.1 & %
			32 & %
			31.4 & %
			0 & %
			55 & %
			53.9 & %
			0 & %
			6 \\ %
			
			471.omnetpp &
			1.17 MB & %
			0:04:00 & %
			802 & %
			800 & %
			0 & %
			99.8 & %
			100.0 & %
			245 & %
			30.6 & %
			0 & %
			530 & %
			66.1 & %
			0 & %
			0 \\ %
			
			473.astar &
			0.04 MB & %
			0:00:00 & %
			1 & %
			1 & %
			0 & %
			100.0 & %
			100.0 & %
			0 & %
			0.0 & %
			0 & %
			0 & %
			0.0 & %
			0 & %
			0 \\ %
			
			483.xalancbmk &
			7.17 MB & %
			5:54:25 & %
			13,440 & %
			12,915 & %
			17 & %
			96.1 & %
			99.9 & %
			2,122 & %
			15.8 & %
			0 & %
			3,792 & %
			28.2 & %
			1 & %
			0 \\ %
			
			\cmidrule(lr){1-15}
			\multicolumn{6}{@{}l}{\emph{Average [SPEC CPU2006]}} &
			\emph{97.8} & %
			\emph{95.6} & %
			& %
			\emph{20.4} & %
			& %
			& %
			\emph{35.1} & %
			& %
			\\[1ex] %
			
			\midrule
			
			510.parest\_r &
			12.69 MB & %
			1:00:00 & %
			4,678 & %
			4,288 & %
			528 & %
			91.7 & %
			89.0 & %
			660 & %
			14.1 & %
			13 & %
			(660) & %
			(14.1) & %
			(13) & %
			-- \\ %
			
			511.povray\_r &
			1.20 MB & %
			0:00:05 & %
			122 & %
			122 & %
			14 & %
			100.0 & %
			89.7 & %
			33 & %
			27.1 & %
			0 & %
			62 & %
			50.8 & %
			0 & %
			6 \\ %
			
			520.omnetpp\_r &
			3.60 MB & %
			0:06:57 & %
			6,430 & %
			6,190 & %
			23 & %
			96.3 & %
			99.6 & %
			1,585 & %
			24.7 & %
			0 & %
			2,286 & %
			35.6 & %
			6 & %
			0 \\ %
			
			523.xalancbmk\_r &
			10.34 MB & %
			15:20:40 & %
			33,880 & %
			33,069 & %
			12 & %
			97.6 & %
			100.0 & %
			1,948 & %
			5.8 & %
			0 & %
			4,961 & %
			14.6 & %
			0 & %
			0 \\ %
			
			526.blender\_r &
			11.47 MB & %
			0:03:29 & %
			174 & %
			172 & %
			80 & %
			98.9 & %
			68.3 & %
			66 & %
			37.9 & %
			0 & %
			70 & %
			40.2 & %
			0 & %
			49 \\ %
			
			541.leela\_r &
			0.33 MB & %
			0:00:01 & %
			1 & %
			1 & %
			0 & %
			100.0 & %
			100.0 & %
			0 & %
			0.0 & %
			0 & %
			0 & %
			0.0 & %
			0 & %
			0 \\ %
			
			\cmidrule(lr){1-15}
			\multicolumn{6}{@{}l}{\emph{Average [SPEC CPU2017]}} &
			\emph{97.4} & %
			\emph{91.1} & %
			& %
			\emph{18.3} & %
			& %
			& %
			\emph{25.9} & %
			& %
			\\[1ex] %
			
			\midrule
			
			MongoDB &
			48.22 MB & %
			1:57:39 & %
			17,836 & %
			16,366 & %
			44 & %
			91.8 & %
			99.7 & %
			552 & %
			3.1 & %
			0 & %
			(552) & %
			(3.1) & %
			(0) & %
			-- \\ %
			
			MySQL &
			35.95 MB & %
			65:57:27 & %
			11,876 & %
			11,592 & %
			179 & %
			97.6 & %
			98.5 & %
			1,330 & %
			11.2 & %
			3 & %
			(1,330) & %
			(11.2) & %
			(3) & %
			-- \\ %
			
			Node.js &
			38.13 MB & %
			5:16:09 & %
			12,643 & %
			12,330 & %
			353 & %
			97.5 & %
			97.2 & %
			1,538 & %
			12.2 & %
			10 & %
			2,559 & %
			20.2 & %
			45 & %
			118 \\ %
			
			\bottomrule
		\end{tabular}
	}
	\label{tab:vtv}
\end{table*}

\begin{table}
	\centering
	\caption{Results of our comparison against CFIXX. For each application this table
		shows (i) the ground truth generated by CFIXX; (ii) static vcall identification,
		depicting the number of indirect call instructions identified as vcall that
		are true positives, the false positives, recall and precision.}
	\scalebox{0.9} {
		\begin{tabular}{l r r r r r }
			\toprule
			
			&
			& \multicolumn{4}{c}{\tabh{Static Identification}} \\
			\cmidrule(lr){3-6}
			
			\tabh{Program} &
			\tabshc{GT} &
			\tabshc{TP} &
			\tabshc{FP} &
			\tabsh{Recall (\%)} &
			\tabsh{Precision (\%)}
			\\ \midrule
			
			447.dealII &
			-- & %
			-- & %
			-- & %
			-- & %
			-- \\ %
			
			450.soplex &
			553 & %
			553 & %
			10 & %
			100.0 & %
			98.2 \\ %
			
			453.povray &
			110 & %
			110 & %
			11 & %
			100.0 & %
			90.9 \\ %
			
			471.omnetpp &
			943 & %
			942 & %
			0 & %
			99.9 & %
			100.0 \\ %
			
			473.astar &
			1 & %
			1 & %
			0 & %
			100.0 & %
			100.0 \\ %
			
			483.xalancbmk &
			12,670 & %
			12,427 & %
			527 & %
			98.0 & %
			95.9 \\ %
			\cmidrule(lr){1-6}
			\multicolumn{4}{@{}l}{\emph{Average [SPEC CPU2006]}} &
			99.6 & %
			97.0 \\[1ex] %
			
			\midrule
			
			510.parest\_r &
			7,288 & %
			7,194 & %
			265 & %
			98.7 & %
			96.5 \\ %
			
			511.povray\_r &
			119 & %
			119 & %
			11 & %
			100.0 & %
			91.5 \\ %
			
			520.omnetpp\_r &
			6,037 & %
			6,032 & %
			71 & %
			99.9 & %
			98.8 \\ %
			
			523.xalancbmk\_r &
			23,661 & %
			26,407 & %
			528 & %
			98.9 & %
			97.8 \\ %
			
			526.blender\_r &
			-- & %
			-- & %
			-- & %
			-- & %
			-- \\ %
			
			541.leela\_r &
			2 & %
			2 & %
			0 & %
			100.0 & %
			100.0 \\ %
			
			\cmidrule(lr){1-6}
			
			\multicolumn{4}{@{}l}{\emph{Average [SPEC CPU2017]}} &
			99.5 & %
			96.9 \\[1ex] %
			
			\midrule
			
			MongoDB &
			20,873 & %
			20,716 & %
			448 & %
			99.3 & %
			97.9 \\ %
			
			MySQL &
			13,035 & %
			12,921 & %
			380 & %
			99.1 & %
			97.1 \\ %
			
			Node.js &
			13,013 & %
			12,982 & %
			491 & %
			99.8 & %
			96.4 \\ %
			
			\bottomrule
		\end{tabular}
	}
	\label{tab:cfixx}
\end{table}

In order to measure the accuracy of the protection of \approachname,
we evaluate the accuracy of the vcall identification analysis.
The results show that \approachname, although a binary-only approach, 
can almost reach the same degree of protection as a source based approach.
Compared to the state-of-the-art binary-only approach \emph{Marx}, it identifies
more vcalls with fewer false-positives.
As applications for our evaluation, we use the C++ programs of SPEC CPU2006
and SPEC CPU2017 that contain virtual callsites, as well as the MySQL server binary (5.7.21),
the Node.js binary (8.10.0), and the MongoDB binary (3.2.4). We used the default optimization levels
(\texttt{O2} for CPU 2006, \texttt{O3} for all others).
The analysis was performed on Ubuntu 16.04 LTS running on an Intel Core i7-2600 CPU with 32 GB of RAM.

\paragraph{VTV}

\begin{figure}
	\small
	\centering
	\begin{subfigure}{\columnwidth}
		\begin{lstlisting}[language=newton,
		numbers=left,
		frame=none,
		xleftmargin=5mm,
		showlines=true,
		firstnumber=92]
		/**
		* Destroy the object pointed to by a pointer type.
		*/
		template<typename _Tp>
		    inline void
		    _Destroy(_Tp* __pointer)
		    { __pointer->~_Tp(); }
		\end{lstlisting}
		\vspace{2mm}
		\caption{Snippet from \texttt{stl\_construct.h}.\label{lst:stlconstruct}}
	\end{subfigure}
	
	\vspace*{3mm}
	\begin{subfigure}{\columnwidth}
		\begin{lstlisting}[language=newton,
		numbers=left,
		frame=none,
		xleftmargin=5mm,
		showlines=true,
		firstnumber=2545]
		Vector<double> us[dim];
		for (unsigned int i=0; i<dim; ++i)
		    us[i].reinit (dof_handler.n_dofs());
		\end{lstlisting}
		\vspace{2mm}
		\caption{Snippet from \texttt{grid\_generator.cc}.\label{lst:gridgenerator}}	
	\end{subfigure}
	\caption{Two source code snippets where VTV fails to identify a virtual callsite.\label{vtverrors}}
\end{figure}

In order to gain a ground truth of
virtual callsites, we use \emph{VTV}~\cite{tice2014enforcing} and compare against
our analysis results. Since VTV leverages source code
information, its results are usually used as ground truth for
binary-only approaches focusing on C++ virtual callsites.
All programs except MongoDB are compiled with GCC 8.1.0. MongoDB crashed during compilation
and had to be compiled with the older version GCC 4.9.3. 
Unfortunately, compiling \emph{450.soplex} results in a crash
and it is therefore omitted. Table~\ref{tab:vtv} shows the
results of our vcall accuracy evaluation.

Overall, we observe that the analysis of \approachname is capable of identifying
the vast majority of virtual callsites in the binary.
This ranges from 91.7\% (\emph{510.parest\_r}) to all vcalls detected (several benchmarks).
Our average recall is 97.8\% on SPEC CPU2006 and 97.4\% on SPEC CPU2017.
With the exception of one outlier (\emph{526.blender\_r} with precision 68.3\%)
we have a low number of false positives, with precision ranging from 87.0\% (\emph{447.dealII})
to no false positives at all (several benchmarks).
The results are similar for large real-world applications with a recall ranging from 91.8\%
(\emph{MongoDB}) to 97.6\% (\emph{MySQL}) and a precision ranging from 97.2\% (\emph{Node.js})
to 99.7\% (\emph{MongoDB}).
The high recall rate shows that our binary-only approach is able to protect almost as many
virtual callsites as VTV does and hence provides comparable security as this source based approach.
However, it still misses some vcalls which may leave an attacker with a small room to perform an attack
under the right circumstances. The precision rates show that although we have a low false positive
identification rate, we still have some.

In order to cope with the problem of false positive identifications, we verify 
vcalls before we actually instrument them with our security check.
The static analysis verification is able to verify 37.9\% in the best case
(\emph{526.blender\_r}) and in the worst case none.
On average we verified 20.4\% on SPEC CPU2006 and 18.3\% on SPEC CPU2017.
For large applications, the best verification rate is 12.2\% (\emph{Node.js}) and the worst
3.1\% (\emph{MongoDB}).
Dynamic verification (see Section~\ref{sec:dynamic_profiling}) considerably
improves verification performance, verifying 35.1\% and 25.9\% for SPEC CPU2006 and 2017.
Unfortunately, we were not able to execute \emph{510.parest\_r}, \emph{MySQL} and \emph{MongoDB}
with VTV. The applications crashed with an error message stating that VTV was
unable to verify a vtable pointer (i.e., a false positive).
Hence, the only large real-world application with dynamic verification \emph{Node.js} verified 20.2\%
of the vcalls.

A manual analysis of the missed virtual callsites (false negatives) reveals two possibilities for
a miss: the data flow was too complex to be handled correctly by our implementation,
or the described pattern in Section~\ref{sec:vcall_candidates} was not used.
The former can be fixed by improving the implemented algorithm that is used
for finding the described pattern. 
In the latter, the \vtblptr is extracted from the
object, however, a newly created stack object is used as \thisptr for
the virtual callsite which does not follow a typical C++ callsite pattern.
This could be addressed by considering additional vcall patterns,
at the risk of adding false positives. Given our already high recall rates,
we believe this would not be a favorable trade-off.

We also verified 86 cases which
VTV did not recognize as virtual callsite instructions. A manual verification of
all cases show that these are indeed vcall instructions and hence
missed virtual callsites by VTV.
For example, Figure~\ref{lst:stlconstruct} depicts the relevant code for 34 of
these cases that are linked to the compiler provided file \texttt{stl\_construct.h}.
Line 98 provides the missed vcall instruction that calls the destructor of the provided
object. Since the destructor of a class is also a virtual function, it is invoked
with the help of a virtual callsite.
Another example is given in Figure~\ref{lst:gridgenerator} for \emph{510.parest\_r}.
Here a vector is created and the function \texttt{reinit()} is invoked on line 2547.
However, since the class \texttt{dealii::Vector<double>} is provided by the
application and \texttt{reinit()} is a virtual function of this class,
this function call is translated into a virtual callsite.
We contacted the VTV authors about this issue and they confirmed that this
happens because the compiler accesses the memory of the objects directly
when calling the virtual function in the internal intermediate representation.
Usually, the compiler accesses them while going through an internal \vtblptr field.
Unfortunately, to fix this issue in VTV would require a lot of non-trivial work since
the analysis has to be enhanced.

\paragraph{CFIXX}

Since \emph{CFIXX} performs the enforcement in a similar way, we also evaluated our
binary-only approach against this source code based method. Hence, we compiled
the applications with CFIXX which is based on LLVM and extracted the protected virtual callsites as
ground truth for our comparison. Table~\ref{tab:cfixx} shows the results of this evaluation.
Unfortunately, we were not able to compile
\emph{447.dealII} and \emph{526.blender\_r} with CFIXX.
As the table shows, \approachname can identify on average 99.6\% of
all SPEC CPU2006 and 99.5\% of SPEC CPU2017 virtual callsites that are also protected
by CFIXX. Furthermore, \approachname also yields a high precision with 97.0\% for SPEC CPU2006
and 96.9\% for SPEC CPU2017 on average. For large real-world applications, the recall and precision
rates are similar with a recall of 99.1\% for \emph{MySQL} and 99.8\% for \emph{Node.js} and
a precision of 97.1\% and 96.4\% respectively.
A manual analysis of the missed virtual callsites (false negatives) showed the same two reasons for a miss
that also occurred for VTV.

\begin{table}
	\centering
	\caption{Results of \emph{Marx}'s vcall accuracy evaluation. For each application this table
		shows (i) the ground truth generated by VTV; (ii) static vcall identification,
		depicting the number of indirect call instructions identified as vcall that
		are true positives, the false positives, recall and precision.}
	\scalebox{0.9} {
		\begin{tabular}{l r r r r r }
			\toprule
			
			&
			& \multicolumn{4}{c}{\tabh{Static Identification}} \\
			\cmidrule(lr){3-6}
			
			\tabh{Program} &
			\tabshc{GT} &
			\tabshc{TP} &
			\tabshc{FP} &
			\tabsh{Recall (\%)} &
			\tabsh{Precision (\%)}
			\\ \midrule
			
			447.dealII &
			1,558 & %
			1,307 & %
			122 & %
			83.9 & %
			91.5 \\ %
			
			450.soplex &
			-- & %
			-- & %
			-- & %
			-- & %
			-- \\ %
			
			453.povray &
			102 & %
			98 & %
			10 & %
			96.1 & %
			90.7 \\ %
			
			471.omnetpp &
			802 & %
			701 & %
			3 & %
			87.4 & %
			99.6 \\ %
			
			473.astar &
			1 & %
			1 & %
			0 & %
			100.0 & %
			100.0 \\ %
			
			483.xalancbmk &
			-- & %
			-- & %
			-- & %
			-- & %
			-- \\ %
			\cmidrule(lr){1-6}
			\multicolumn{4}{@{}l}{\emph{Average [SPEC CPU2006]}} &
			91.8 & %
			95.4 \\[1ex] %
			
			\midrule
			
			510.parest\_r &
			4,678 & %
			3,673 & %
			295 & %
			78.5 & %
			92.6 \\ %
			
			511.povray\_r &
			122 & %
			115 & %
			11 & %
			94.3 & %
			91.3 \\ %
			
			520.omnetpp\_r &
			6,430 & %
			5,465 & %
			22 & %
			85.0 & %
			99.6 \\ %
			
			523.xalancbmk\_r &
			33,880 & %
			23,541 & %
			33 & %
			69.4 & %
			99.9 \\ %
			
			526.blender\_r &
			174 & %
			171 & %
			1,347 & %
			98.3 & %
			11.3 \\ %
			
			541.leela\_r &
			1 & %
			0 & %
			0 & %
			0.0 & %
			0.0 \\ %
			
			\cmidrule(lr){1-6}
			
			\multicolumn{4}{@{}l}{\emph{Average [SPEC CPU2017]}} &
			70.9 & %
			65.8 \\[1ex] %
			
			\midrule
			
			MongoDB &
			17,836 & %
			12,437 & %
			1,249 & %
			69.7 & %
			90.9 \\ %
			
			MySQL &
			11,876 & %
			10,867 & %
			1,214 & %
			81.3 & %
			88.8 \\ %
			
			Node.js &
			12,643 & %
			10,648 & %
			1,095 & %
			84.2 & %
			90.7 \\ %
			
			\bottomrule
		\end{tabular}
	}
	\label{tab:marx}
\end{table}

\paragraph{Marx}

A direct comparison of the accuracy with other binary-only approaches is
difficult since different test sets are used to evaluate it.
For example, vfGuard evaluates the accuracy of their approach against
only two applications, while T-VIP is only evaluated against one.
VTint states absolute numbers without any
comparison with a ground truth. VCI evaluates their approach against
SPEC CPU2006, but the numbers given for the ground truth created with VTV
differ completely from ours (\eg, 9,201 vs. 13,440 vcalls for \emph{483.xalancbmk})
which makes a comparison difficult. Additionally, the paper reports
no false positives during their analysis which we encounter in the same
application set with a similar identification technique. Unfortunately,
as discussed in Section~\ref{sec:vcall_candidates},
we were not able to determine the reason for this.
Furthermore, most approaches target different platforms
than \approachname (Windows x86 and Linux x86) and are not open source.
Since Marx is the only open source approach that targets the same platform, we analyzed our evaluation set with it.
In order to create as few false positives as possible we used its conservative
mode. Unfortunately, Marx crashed during the analysis of
\emph{483.xalancbmk}. The results of the analysis can be
seen in Table~\ref{tab:marx}.
Compared to Marx, we have considerably higher recall with better precision.
Averaged over the CPU2006 benchmarks supported by Marx,
\approachname achieves 98.2\% recall (91.8\% for Marx) and
on CPU2017 97.4\% versus 70.9\%, respectively.
This does not come at the cost of more false positives,
as our precision is similar on CPU2006 (94.5\% vs. 95.4\%)
and much better on CPU2017 (91.1\% vs. 65.8\%).
For large real-world applications like \emph{MySQL} and \emph{MongoDB},
\approachname identifies 16.3\% and 28.1\% more virtual callsites
with better precision (98.5\% vs. 88.8\% for \emph{MySQL} and
99.7\% vs. 90.9\% for \emph{MongoDB}).
\medskip

Overall, our analysis shows that \approachname is precise enough to provide an application
with protection against control-flow hijacking attacks at virtual callsites.
The evaluation showed that on average only 2.5\% when comparing against VTV and
0.5\% comparing against CFIXX of the vcalls were missed.
Since binary analysis is a hard problem, the results are very promising in showing
that a sophisticated analysis can almost reach the same degree of protection as a source based approach.
In addition, it shows that even source code approaches such as VTV do not
find all virtual callsite instructions and can benefit from binary-only 
approaches such as \approachname. Furthermore, the number of
false positives shows the sensibility of our approach to handle
them during instrumentation rather than assume their absence.

\subsection{Object Initialization/Destruction Accuracy}

\begin{table}
	\centering
	\caption{Object creation and destruction accuracy results, showing the number of vtable references in the code as found in the
		ground truth, and as identified or missed by our analysis.}
	\scalebox{0.9} {
		\begin{tabular}{l r r r }
			\toprule
			
			\tabh{Program} &
			\tabshc{GT} &
			\tabshc{identified} &
			\tabshc{missed}
			\\ \midrule
			
			447.dealII &
			-- &
			-- &
			-- \\ %
			
			450.soplex &
			102 &
			228 &
			0 \\ %
			
			453.povray &
			103 &
			226 &
			0 \\ %
			
			471.omnetpp &
			372 &
			871 &
			0 \\ %
			
			473.astar &
			0 &
			8 &
			0 \\ %
			
			483.xalancbmk &
			2,918 &
			6,530 &
			0 \\ %
			
			\midrule
			
			510.parest\_r &
			12,482 &
			25,804 &
			0 \\ %
			
			511.povray\_r &
			103 &
			224 &
			0 \\ %
			
			520.omnetpp\_r &
			1,381 &
			3,280 &
			0 \\ %
			
			523.xalancbmk\_r &
			2,790 &
			6,323 &
			0 \\ %
			
			526.blender\_r &
			-- &
			-- &
			-- \\ %
			
			541.leela\_r &
			87 &
			180 &
			0 \\ %
			
			\midrule
			
			MongoDB &
			8,054 & %
			11,401 & %
			0 \\ %
			
			MySQL&
			8,532 &
			11,524 &
			0 \\ %
			
			Node.js &
			7,816 &
			19,204 &
			0 \\ %
			
			\bottomrule
		\end{tabular}
	}
	
	\label{tab:objinit}
\end{table}

To avoid breaking applications, \approachname must instrument all valid object initialization and destruction sites.
To ensure that this is the case, we compare the number of vtable-referencing instructions found by \approachname to a ground truth.
We generate the ground truth with an LLVM 4.0.0 pass
that instruments Clang's internal function
\texttt{CodeGenFunction::InitializeVTablePointer()}, which Clang uses for all vtable pointer initialization.

Table~\ref{tab:objinit} shows the results for the same set of applications
we used in Section~\ref{sec:evalvcalls}. We omit
results for \emph{447.dealII} from SPEC CPU 2006 and
\emph{526.blender\_r} from SPEC CPU 2017 because these benchmarks fail to compile with LLVM 4.0.0.
The results for the remaining applications show that our analysis finds all vtable-referencing instructions.
It conservatively overestimates the set of vtable-referencing
instructions, ensuring the security and correctness of \approachname at the cost of a slight performance degradation due to the overestimated instruction set.

\subsection{Performance}

\begin{table*}
	\centering
	\caption{\approachname performance results and runtime statistics. For each binary, this table shows \first
		\textbf{binary instrumentation} details, depicting the number of instrumented
		\vtblptr writes (\textit{\#vtblptr}), positive virtual calls
		(\textit{\#positive}), and candidate
		vcalls (\textit{\#candidates}); \second \textbf{runtime statistics}, listing
		the number of true positive (\textit{\#TP}) and false positive
		(\textit{\#FP}) virtual calls,
		and the total number of virtual calls (\textit{\#vcalls}) and 
		\vtblptr writes (\textit{\#vtblptr});
		and \third \textbf{runtime overhead}, listing runtime overhead
		(\textit{\approachname}) compared to the baseline (\textit{base}) in seconds. 
		\label{tab:spec2017-stats}}
	\begin{tabular*}{\textwidth}{@{\extracolsep{\fill}}l r r r r r r r r r} 
\toprule

  & \multicolumn{3}{c}{\tabh{Binary instrumentation}} 
  & \multicolumn{4}{c}{\tabh{Runtime statistics}}
  & \multicolumn{2}{c}{\tabh{Runtime overhead}} \\

    \cmidrule(lr){2-4}
    \cmidrule(lr){5-8}
    \cmidrule(lr){9-10}

    & \tabshc{vtblptr} & \tabshc{positive} & \tabshc{candidates} 
    & \tabshc{TP} & \tabshc{FP} & \tabshc{vcalls} & \tabshc{vtblptr} 
    & \tabsh{base} & \tabsh{\vps}\phantom{ (+00\%)}\\

\midrule
    444.namd      &      6 &     0 &      2 &     0 &  0 &             0 &         2,018 & 343.5 & 342.9 (+\phantom{0}0\%) \\
    447.dealII    &  4,283 &   161 &  1,459 &    47 &  0 &           97m &           21m & 289.7 & 299.2 (+\phantom{0}3\%) \\
    450.soplex    &    120 &   195 &    364 &    48 &  0 &     1,665,968 &            40 & 215.8 & 220.2 (+\phantom{0}2\%) \\
    453.povray    &     98 &    21 &     91 &    21 &  6 &       101,743 &           162 & 135.8 & 153.3 (+13\%) \\
    471.omnetpp   &    507 &   117 &    677 &   327 &  0 &        1,585m &        2,156m & 290.0 & 370.2 (+28\%) \\
    473.astar     &      0 &     0 &      1 &     0 &  0 &             0 &             0 & 350.3 & 351.6 (+\phantom{0}0\%) \\
    483.xalancbmk &  4,554 & 1,348 & 11,623 & 1,639 &  0 &        3,822m &        2,316m & 185.0 & 249.4 (+35\%) \\
    \cmidrule(lr){1-10}
    
    \multicolumn{9}{@{}l}{\emph{Geometric mean [SPEC CPU2006]}} & +\phantom{(}11\%\phantom{)} \\[1ex] 

    \midrule                                                                                  
    508.namd\_r      &     48 &     0 &      0 &     0 &  0 &             0 &            21 & 271.8 & 271.8 (+\phantom{0}0\%) \\
    510.parest\_r    & 12,206 &   243 &  4,539 &   350 &  4 &        2,625m &          119m & 586.3 & 603.1 (+\phantom{0}3\%) \\
    511.povray\_r    &    113 &    19 &    121 &    21 &  6 &         4,577 &           183 & 498.7 & 572.0 (+15\%)           \\
    520.omnetpp\_r   &  2,591 &   447 &  5,310 &   751 &  0 &        7,958m &        2,070m & 507.4 & 661.7 (+30\%)           \\
    523.xalancbmk\_r &  4,512 &   801 & 30,771 & 2,844 &  0 &        4,873m &        2,314m & 366.8 & 461.5 (+26\%)           \\
    526.blender\_r   &     43 &    37 &    174 &     4 & 46 &            11 &             3 & 325.8 & 328.6 (+\phantom{0}1\%) \\
    531.deepsjeng\_r &      0 &     0 &      0 &     0 &  0 &             0 &             0 & 345.1 & 353.1 (+\phantom{0}2\%) \\
    541.leela\_r     &    177 &     0 &      2 &     0 &  0 &             0 &       404,208 & 535.5 & 534.6 (+\phantom{0}0\%) \\
    \cmidrule(lr){1-10}

    \multicolumn{9}{@{}l}{\emph{Geometric mean [SPEC CPU2017]}} & +\phantom{(0}9\%\phantom{)} \\

\bottomrule
\end{tabular*}

\end{table*}

\begin{figure}
	\small
	\centering
	\begin{subfigure}{\columnwidth}
		\centering
		\resizebox{\columnwidth}{!}{\input{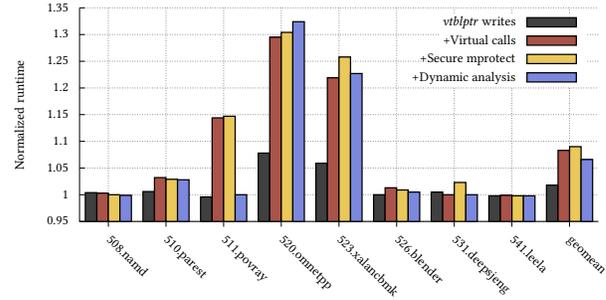}}
		\vspace{-9mm}
		\caption{Microbenchmarks for SPEC CPU2017\label{fig:spec2017}}
	\end{subfigure}
	\begin{subfigure}{\columnwidth}
		\centering
		\resizebox{\columnwidth}{!}{\input{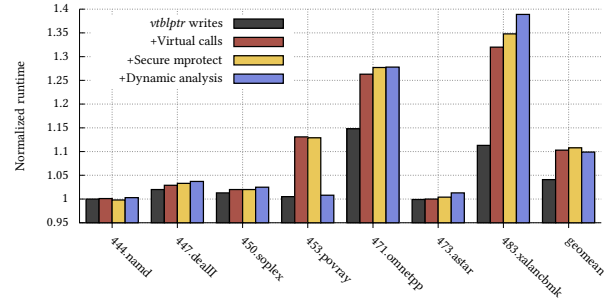}}
		\vspace{-9mm}
		\caption{Microbenchmarks for SPEC CPU2006\label{fig:spec2006}}
	\end{subfigure}
	\caption{Normalized runtime for C++ programs in SPEC CPU2006 and
		CPU2017, with cumulative configurations: \first only instrument \vtblptr \emph{writes};
		\second also instrument \emph{virtual call instructions};
		\third \emph{secure the
			safe region} by marking all pages unwritable, and only selectively
		\texttt{mprotect}-ing them if they are accessed from our own
		instrumentation code; and \fourth include offline \emph{dynamic analysis} results,
		reducing the need for hot-patching.
		\label{fig:spec2017bars}}
\end{figure}

This section evaluates
the runtime performance of \approachname by measuring the time it takes to run each 
C++ benchmark in SPEC CPU2006 and CPU2017.
We compare \approachname-protected runtimes against the baseline of original
benchmarks without any instrumentation.
We compile all test cases as position-indepedent executables with GCC 6.3.0.
For each benchmark, we report the median runtime over 11 runs on a Xeon E5-2630
with 64\,GB RAM, running CentOS Linux 7.4 64-bit. We use a single additional
run with more logging enabled to obtain statistics such as
the number of executed virtual calls. Table~\ref{tab:spec2017-stats}
details our results.

Our results show the variety in properties of C++ applications. Some programs make
little to no use of virtual dispatching, \eg, \emph{444.namd}, \emph{508.namd\_r},
\emph{531.deepsjeng\_r}, and \emph{473.astar}.
Others contain thousands of \vtblptr writes and virtual callsites, \eg, \emph{510.parest\_r}
with over 12,000 \vtblptr writes, or \emph{483.xalancbmk} in CPU2006 with more than 1,300
verified virtual callsites. Further details are shown in the first group in Table~\ref{tab:spec2017-stats}.

The comparison of verified virtual calls (true positive) and regular indirect calls (false positive)
shows the accuracy of our analysis. Almost all vcall candidates turn out to be real vcalls.
Furthermore, with absolute numbers of executed virtual calls and \vtblptr writes
in the billions, it is clear that our instrumentation must be lightweight.
The second group in Table~\ref{tab:spec2017-stats} depicts the exact numbers.

The runtime overhead of our instrumentation varies from 0\% for programs with little to no
virtual dispatch code to 35\% for the worst-case scenario (\emph{483.xalancbmk}). In
almost all cases, we see a correlation between increased overhead and number
of instrumentation points (\vtblptr writes and virtual calls). An exception
is \emph{511.povray\_r}, which shows a 15\% performance decrease despite a relatively
low number of vcalls and \vtblptr writes. Further
inspection shows that this is caused by the 6 false positives candidate
vcalls; if we disable hot-patching, our vcall instrumentation code is called
over 18 billion times. While we remove instrumentation hooks for the majority of these cases,
which are not real vcalls, our current implementation does not remove the
Dyninst trampolines. These trampolines are the source of the unexpected overhead.
The numbers depicting the comparison of the uninstrumented baseline runs to
\approachname-protected runs are shown in the third group in Table~\ref{tab:spec2017-stats}.

To better understand the overhead of \approachname, we gathered detailed statistics for both SPEC
CPU2006 and SPEC CPU2017 in varying configurations. 
We first run SPEC with only instrumentation for \vtblptr writes
enabled. In this run, the entire safe region is 
read/writable and the instrumentation only \first computes the
address in the safe region to store the vtable pointer at, and
\second copies the vtable pointer there.  In the second configuration,
we additionally instrument virtual calls. We check whether candidates
are actual vcalls by testing the call's first argument
and, if it can be dereferenced, looking this value up in the list of known vtables. 
We then either patch verified
vcalls to enable the fast path, or remove instrumentation for false positives.
The fast path 
fetches the vtable pointer by dereferencing the first argument, and then
compares it against the value stored in the safe region. The third
configuration additionally makes the safe region read-only and uses a segfault handler to mark
pages writable on demand. Finally, the fourth configuration
includes dynamic analysis results, removing the need to hot-patch
previously verified vcalls at runtime.
The results show that the majority of \approachname's overhead
stems from \first \vtblptr writes, and \second 
virtual callsite instrumentation. Figure~\ref{fig:spec2017bars}
details the numbers of this evaluation.

Overall, with a geometric mean performance overhead of 11\% for SPEC CPU2006 and 9\% for SPEC CPU2017,
\approachname shows a moderate performance impact.
As expected, it does not perform as well as a source-based approach such as VTV with reported 4\% geometric
mean for SPEC CPU2006~\cite{tice2014enforcing}. However, it outperforms comparable previous work
(VCI with 14\% \cite{elsabagh2017strict} and T-VIP with 25\%~\cite{gawlik2014towards}) and performs slightly worse
than Marx's \emph{VTable Protection} with a reported 8\% geometric mean for SPEC CPU2006, however, with
better accuracy and additional type integrity.

\section{Discussion}

This section first discusses the susceptibility of \approachname to 
Counterfeit Object-oriented Programming~\cite{schuster2015counterfeit}.
Following this, we discuss the limitations of \approachname.

\subsection{Counterfeit Object-oriented Programming}

CFI approaches targeting C++ must cope with
advanced attackers using Counterfeit Object-oriented Programming
(\emph{COOP}) attacks \cite{schuster2015counterfeit,crane2015s}.
This attack class thwarts defenses that do not accurately model C++ semantics.
As we argue below, \approachname reduces the attack surface sufficiently
that practical COOP attacks are infeasible.

For a successful \emph{COOP} attack, an attacker must control a
container filled with objects, with a loop invoking a virtual function on each object.
The loop may be an actual loop, called a \emph{main loop gadget},
or can be achieved through recursion, called a \emph{recursion gadget}.
We refer to both types as \emph{loop gadget}.
The attacker places counterfeit objects in the container, allowing them
to hijack control flow when the loop executes each object's virtual function.
To pass data between the objects, the attacker can
overlap the objects' fields.

The first restriction \approachname imposes on an attacker is that it prevents 
filling the container with counterfeit objects;
because the objects were not created at legitimate object creation sites, the safe memory does not contain stored \vtblptr{}s for them.
Only two conceivable ways would allow an attacker to craft a container of counterfeit objects under \approachname:
either the application allows attackers to arbitrarily invoke constructors and create objects, 
or the attacker can coax the application into creating all objects needed for an attack through legitimate behavior.
The former occurs (in restricted form) only in applications with scripting capabilities.
The latter scenario, besides requiring an cooperative victim application, hinges on the attacker's ability to scan data memory to find all needed objects without crashing the application (hence losing the created objects)
and filling the container with pointers to these.

The second restriction \approachname imposes is that it prohibits overlapping objects (used for data transfer in COOP) because objects can only be created through legitimate constructors.
This means that a would-be COOP attack would have to pass data via argument registers or via a scratch memory area instead.
Data passing via argument registers works only if the loop gadget does not modify the argument registers between gadget invocations.
Additionally, the virtual functions used as gadgets must leave their results in precisely the correct argument registers when they return.
Passing data via scratch memory limits the attack to the use of virtual functions
that work on memory areas. The pointer to the scratch memory area must then be passed
to the virtual function gadgets either via an argument register (subject to the limitations
of passing data via argument registers), or via a field in the object.
To use a field
in the object as a pointer to scratch memory, the attacker must overwrite
that field prior to the attack, which could lead to a crash if the application tries to use the modified object.

As a third restriction, \approachname's checks of the \vtblptr at each vcall instruction
mean that the attacker is limited in the virtual functions they can use
at a \emph{loop gadget}. Only the virtual function at the specific vtable offset used by the vcall
is allowed; attackers cannot ``shift'' vtables to invoke alternative entries. This security policy is comparable to
\emph{vfGuard}~\cite{prakash2015vfguard}.

To summarize, \approachname restricts three crucial COOP components: object creation, data transfer, and
\emph{loop gadget} selection.
Because all proof-of-concept exploits by
Schuster~et~al.~\cite{schuster2015counterfeit} rely on object overlapping as
a means of transferring data, \approachname successfully prevents them.
Moreover, Schuster~et~al. recognize \emph{vfGuard} as a significant constraint
for an attacker performing a \emph{COOP} attack.
Given that \approachname raises the bar even more than \emph{vfGuard}, we argue that \approachname makes currently existing
COOP attacks infeasible.

We found that multiple of the virtual callsites missed by \emph{VTV} (as shown in Section~\ref{sec:evalvcalls}) reside in a loop
in a destructor function (similar to the \emph{main loop gadget} example used
by Schuster~et~al.~\cite{schuster2015counterfeit}). Because the loop iterates
over a container of objects and uses a virtual call on each object, COOP attacks can leverage these missed callsites as a \emph{main loop gadget} even with \emph{VTV} enabled.
This demonstrates the need for defense-in-depth, with multiple hurdles for an attacker to cross in case of inaccuracies in the analysis.

\subsection{Limitations}
\label{sec:limitations}

At the moment, our proof-of-concept implementation of the instrumentation ignores object deletion
because it does not affect the consistency of the safe memory. 
As a result, when an object is deleted, its old \vtblptr is still
stored in safe memory. If an attacker manages to control
the memory of the deleted object, they can craft a new
object that uses the same vtable as the original object.
Because the \vtblptr remains unchanged,
this attack is analogous to corrupting an object's fields and does not allow the attacker to hijack control.
Thus, while our approach does not completely prevent
use-after-free, it forces an attacker to re-use the type of the object previously stored in the attacked memory.

Another limitation of our approach lies in the runtime verification of candidate vcall sites.
If an attacker uses an unverified vcall instruction,
they can force the analysis instrumentation to detect a ``false positive''
vcall and remove the security instrumentation for this instruction, leaving the vcall unprotected.
Because we cache analysis results, this attack only works for vcall sites that are unverified in the static analysis and have never been executed before in any run of the program (since otherwise only the security check is performed), leading to a race condition between the analysis instrumentation and the attacker.
The only way to mitigate this issue is by improving coverage
during the dynamic profiling analysis and therefore reducing the
number of unverified vcalls.
This is possible by running test cases for the protected program or through techniques such as fuzzing~\cite{triforce,rawat2017vuzzer}.
Note also that this attack requires specific knowledge of an unverified vcall; if the attacker guesses wrong and attacks a known vcall, we detect and log the attack.

\approachname inherits some limitations from Dyninst, such as Dyninst's inability to
instrument functions
that catch or throw C++ exceptions, and Dyninst's inability to instrument
functions for which it fails to reconstruct a CFG. These limitations
are not fundamental to \approachname and can be resolved with additional engineering effort.

Finally, we note that our safe memory region implementation can be enhanced, for example,
by using hardware features such as Memory Protection Keys (MPK)~\cite{mpk}. In the current implementation, an adversary might still
be able to overwrite values in the safe memory region under the right circumstances.
However, because the safe region is merely a building block for \approachname, we consider improvements to safe memory an orthogonal topic and do not explore it further in this work.

\section{Related Work}
\label{sec:related}

\emph{Marx}~\cite{pawlowski2017marx}
reconstructs class hierarchies from binaries
for \emph{VTable Protection}
and \emph{Type-safe Object Reuse}. VTable Protection
verifies at each vcall whether the \vtblptr
resides in the reconstructed class hierarchy.
However, the analysis is incomplete and the instrumentation
falls back to \emph{PathArmor}~\cite{van2015practical} for missing results.
Marx's Type-safe Object Reuse prevents memory reuse between different class hierarchies,
reducing the damage that can be done with use-after-free. However, this approach leaves considerable
wiggle room for attackers for large class hierarchies.
In contrast, \approachname does not rely on class hierarchy information and provides stronger security by only allowing exactly correct types.
Moreover, Marx only protects the heap whereas \approachname
protects all objects.

\emph{VTint}~\cite{zhang2015vtint} instruments vtables with IDs
to check their validity, but unlike \approachname allows exchanging
the original \vtblptr with a new pointer to an existing vtable.
Moreover, VTint breaks the binary in case of false positives.

\emph{VTPin}~\cite{sarbinowski2016vtpin} overwrites the \vtblptr
whenever an object is freed, to protect against use-after-free, but
requires RTTI and does not prevent \vtblptr overwrites in general.

\emph{vfGuard}~\cite{prakash2015vfguard} identifies vtables and
builds a mapping of valid target functions at each vtable
offset. At vcalls, it checks the target and calling convention.
Unlike \approachname, vfGuard allows fake vtables as long as each entry
appears in some valid vtable at the same offset.
Additionally, vfGuard may break the binary in case of false positives.

\emph{T-VIP}~\cite{gawlik2014towards} protects vcalls against
fake vtables,
but breaks the binary when vtables reside
in writable memory (\eg, in \texttt{.bss}).
Moreover, unlike \approachname, T-VIP uses potentially bypassable heuristics.

\emph{VCI}~\cite{elsabagh2017strict} only allows a specific set of vtables
at each vcall, mimicking \emph{VTV}~\cite{tice2014enforcing}.
When the analysis cannot rebuild the sets precisely,
VCI falls back to vfGuard.
Moreover, false positive virtual callsites in VCI
break the application, as may
incomplete class hierarchies (\eg, due to abstract classes~\cite{pawlowski2017marx}).
In contrast, \approachname allows calls through any legitimately created object.
Moreover, even in the hypothetical case of a perfect VCI analysis,
VCI allows changing the \vtblptr to another one
in the set, unlike \approachname.

\emph{VTV}~\cite{tice2014enforcing} is a GCC compiler pass
that only allows a statically determined set of vtables at each vcall, like most binary-only approaches~\cite{elsabagh2017strict,
	gawlik2014towards, prakash2015vfguard, pawlowski2017marx}.

\emph{CFIXX}~\cite{burow2018cfixx} is the state-of-the-art in
source-based C++ defenses. Like \approachname, it stores 
\vtblptr{}s in safe memory.
At each callsite, the \vtblptr is 
fetched from the safe memory region.
Given the lack of comparison against the \vtblptr as stored in the object,
CFIXX prevents but does not detect vtable hijacking.
As an LLVM compiler extension, CFIXX cannot protect applications for which no source code is available. 
Therefore, proprietary legacy applications cannot be protected afterwards.
Moreover, not all software compiles on LLVM out-of-the-box
(e.g., the Linux kernel~\cite{edge2017}).
While CFIXX and \approachname offer similar security,
our binary-level analysis is completely novel.
Unlike source-level analysis, our analysis must consider both direct and indirect
vtable accesses. Moreover, identifying the virtual callsites
for subsequent security instrumentation is challenging given the lack of type information.

\section{Conclusion}

In this paper, we presented \approachname, a practical binary-level defense against C++ vtable hijacking.
While prior work restricts the targets of virtual calls, we protect objects \emph{at creation time} and only allow virtual calls reachable by the object, sidestepping accuracy problems.
\approachname improves correctness by handling false positives at vcall verification.
During our evaluation, we also uncovered several inaccuracies
in \emph{VTV}, a source-based approach that is considered the state-of-the-art for C++ defenses.
We release \approachname as open source software to foster research on this
topic.

\section*{Acknowledgements}

This work was supported by the German Research Foundation (DFG) within the
framework of the Excellence Strategy of the Federal Government and the States --
EXC~2092 \textsc{CaSa} -- 39078197, by the United States Office of Naval
Research under contracts N00014-17-1-2782 and N00014-17-S-B010 ``BinRec'',
and by the European Research Council (ERC) under the European Union's Horizon
2020 research and innovation programme under grant agreement No. 786669 (ReAct),
No. 825377 (UNICORE), and No. 640110 (BASTION). Any opinions,
findings, and conclusions or recommendations expressed in this paper are those
of the authors and do not necessarily reflect the views of any of the sponsors
or any of their affiliates.

\bibliographystyle{ACM-Reference-Format}
\bibliography{paper}

\end{document}